\documentclass[10pt]{article}
\usepackage[cp1251]{inputenc}
\usepackage[dvips]{epsfig}
\usepackage[dvips]{changebar}
\usepackage[dvips]{graphicx}
\usepackage{amsmath,enumerate, amsfonts, amssymb,amsthm}
\usepackage{cite}

\def\lsim{\
  \lower-1.2pt\vbox{\hbox{\rlap{$<$}\lower5pt\vbox{\hbox{$\sim$}}}}\ }
\def\gsim{\
  \lower-1.2pt\vbox{\hbox{\rlap{$>$}\lower5pt\vbox{\hbox{$\sim$}}}}\ }

\begin{document}

\title{Nonuniform Bose-Einstein condensate. II. \\ Doubly coherent states}
\author{Maksim Tomchenko \bigskip \\
{\small Bogolyubov Institute for Theoretical Physics} \\
{\small 14b, Metrolohichna Str., Kyiv 03143, Ukraine}}
\date{\empty }
\maketitle

\textit{We find stationary excited states of a one-dimensional
system of $N$ spinless point bosons with repulsive interaction and
zero boundary conditions by numerically solving the time-independent
Gross-Pitaevskii equation. The solutions are compared with the exact
ones found in the Bethe-ansatz approach. We  show that the $j$th
stationary excited state of a nonuniform condensate of atoms
corresponds to a Bethe-ansatz solution with the quantum numbers
$n_{1}=n_{2}=\ldots =n_{N}=j+1$. On the other hand, such
$n_{1},\ldots,n_{N}$ correspond to a condensate of $N$ elementary
excitations (in the present case the latter are the Bogoliubov
quasiparticles with the quasimomentum $\hbar \pi j/L$, where $L$ is
the system size). Thus, each stationary excited state of the
condensate is \textquotedblleft doubly coherent\textquotedblright,
since it
corresponds simultaneously to a condensate of $N$ atoms and a condensate of $%
N$ elementary excitations. We find the energy $E$ and the particle
density profile $\rho (x)$ for such states. The possibility of
experimental production of these states is also discussed.}

\section{Introduction}

In the previous work \cite{gp1}, the ground state of a nonuniform
condensate was studied. In the present work, we consider the
stationary \textit{excited} states of such a condensate. Excited
states, which are solutions of the general time-dependent
Gross-Pitaevskii
(GP) equation, can be non-stationary (quasiparticles \cite%
{gross1963,tsuzuki1971,stringari1996,edwards1996a,bigelow2003,pethick2008,mtgp2012}
and solitons \cite{tsuzuki1971,pethick2008,zakharov1972, zakharov1973,
agraval2003, malomed2022}) and stationary.\footnotemark\footnotetext{%
Due to an enormous
number of relevant articles, we are
familiar with (and can cite) only some of the theoretical articles;
other references, including experimental articles, can be found in
reviews
\cite{malomed2022,syrwid2021} and monographs \cite%
{pethick2008,agraval2003,leggett2006,pitstring2016}.} The latter are
solutions of a boundary value problem, being given by the stationary
GP equation and boundary conditions (BCs). Such solutions can be of
a
vortex \cite%
{gross1963,ginzburg1958,wu1961,amit,jones,edwards1996v,dalfovo1996,ho1996,jackson1998,burnett1999}
or a non-vortex type. Here we will consider the one-dimensional (1D)
problem, where only non-vortex solutions are possible. Such
solutions have
been found for several systems~\cite%
{edwards1995,carr2000r,carr2000a,kivshar2001}. Seemingly, they were
not observed experimentally.

Below we will find solutions for stationary excited states of a 1D system of
spinless point bosons under zero BCs. Solutions for such a system have
already been obtained and written in terms of the Jacobi elliptic functions~\cite%
{carr2000r}. We will find these solutions differently (numerically)
and compare them with the exact ones obtained by the Bethe ansatz.
Our main aim is to ascertain a relationship between the solutions of
the GP equation and the elementary excitations of the system. To our
knowledge, such an analysis has not been carried out before.

The structure of the paper is as follows. Section~2 contains the
basic equations. The idea about the physical origin of stationary
excited states of the condensate is formulated in section~3. In
sections~4 and 5 we analyze solutions of the time-independent GP
equation corresponding to a hole-like Lieb's quasiparticle and an
ideal crystal. In sections~6 and 7 we find a relationship between
the solutions of the time-independent GP equation, exact solutions
obtained in the Bethe-ansatz approach, and elementary
quasiparticles. The final sections 8 and 9 are devoted to a
discussion of the results obtained and possible experiments. In the
Appendix we investigate whether the condensate $\Phi _{j_{0}}(x)$,
which is a solution of the stationary GP equation, can be
fragmented.

\section{Initial equations}

It is known that the substitution of the $c$-number ansatz%
\begin{equation}
\hat{\Psi}(\mathbf{r},t)=\Psi (\mathbf{r},t)  \label{1}
\end{equation}%
into the Heisenberg equation results in the Gross-Pitaevskii equation~\cite%
{pit1961,gross1961}
\begin{equation}
i\hbar \frac{\partial \Psi (\mathbf{r},t)}{\partial t}=-\frac{\hbar ^{2}}{2m}%
\frac{\partial ^{2}\Psi (\mathbf{r},t)}{\partial \mathbf{r}^{2}}+2c\Psi (%
\mathbf{r},t)|\Psi (\mathbf{r},t)|^{2}.  \label{2}
\end{equation}%
This equation describes a nonuniform condensate in a system of $N$
spinless bosons that interact through a point-like potential $U(|\mathbf{r}%
_{j}-\mathbf{r}_{l}|)=2c\delta (\mathbf{r}_{j}-\mathbf{r}_{l})$. In work
\cite{gp1} it was shown that if, instead of the $c$-number ansatz (\ref{1}%
), we use a somewhat more accurate \textit{operator} ansatz
\begin{equation}
\hat{\Psi}(\mathbf{r},t)=\hat{a}_{0}\Psi (\mathbf{r},t)/\sqrt{N},  \label{3}
\end{equation}%
then the Heisenberg equation leads to the equation
\begin{equation}
i\hbar \frac{\partial \Psi (\mathbf{r},t)}{\partial t}=-\frac{\hbar ^{2}}{2m}%
\frac{\partial ^{2}\Psi (\mathbf{r},t)}{\partial \mathbf{r}^{2}}+\left( 1-%
\frac{1}{N}\right) 2c\Psi (\mathbf{r},t)|\Psi (\mathbf{r},t)|^{2}.  \label{4}
\end{equation}%
This equation also follows from the ansatz
\begin{equation}
\Psi _{N}(\mathbf{r}_{1},\ldots ,\mathbf{r}_{N},t)=\prod\limits_{j=1}^{N}[%
\Psi (\mathbf{r}_{j},t)/\sqrt{N}],  \label{5}
\end{equation}%
as was shown by the analysis of the $N$-particle Schr\"{o}dinger
equation \cite{esryphd} (see also \cite{gp1}) and by a variational
method \cite{salasnich2000}. Below, \textit{Eqs.~(\ref{2}) and
(\ref{4}) will be called the GP and GP$_{N} $ equations,
respectively.} The GP$_{N}$ equation differs from the GP one by the
factor $\left( 1-1/N\right) $.
Ans\"{a}tze~(\ref{1}), (\ref{3}), and (\ref{5}) are equivalent in
the sense
that each of them describes a system where all atoms are in the condensate $%
\Psi (\mathbf{r},t)$. Ansatz (\ref{1}) is valid for $N\gg 1$, and ans\"{a}tze (%
\ref{3}) and (\ref{5}) for $N\geq 2$.

In work \cite{gp1} we studied the solutions of the stationary GP and GP$%
_{N} $ equations, corresponding to the ground state of the
condensate, for different values of $N$, the
average particle density $\bar{\rho}=N/L$, and the coupling constant $%
\gamma =c/\bar{\rho}$. The analysis has shown that the GP$_{N}$
equation describes the system in the near-free particle regime
($\gamma N^{2}\lesssim 1$) much more accurately than the GP equation
does. This regime corresponds to
small $N$  ($N\lesssim \gamma ^{-1/2}$) or ultra-weak coupling ($%
\gamma \lesssim N^{-2}$). For weak coupling ($\gamma \lesssim 0.1$
and $N\gg 1$), the GP and GP$_{N}$ equations describe the system
with the same accuracy when $\gamma N^{2}\gg 1$.

In the present work, we consider a 1D system of $N$ spinless bosons
with point repulsive interaction ($c>0$), which occupy  the segment
$[0,L]$ under zero BCs. We seek stationary solutions
\begin{equation}
\Psi (x,t)=e^{\epsilon t/i\hbar }\Phi (x),  \label{6}
\end{equation}%
so that the GP and GP$_{N}$ equations take the forms
\begin{equation}
\epsilon \Phi (x)=-\frac{\hbar ^{2}}{2m}\frac{\partial ^{2}\Phi }{\partial
x^{2}}+2c|\Phi |^{2}\Phi  \label{9-1}
\end{equation}%
and%
\begin{equation}
\epsilon \Phi (x)=-\frac{\hbar ^{2}}{2m}\frac{\partial ^{2}\Phi }{\partial
x^{2}}+\left( 1-\frac{1}{N}\right) 2c|\Phi |^{2}\Phi ,  \label{9-2}
\end{equation}%
respectively, with the boundary conditions
\begin{equation}
\Phi (x=0)=\Phi (x=L)=0.  \label{8}
\end{equation}

Let us find solutions of Eqs.~(\ref{9-1}) and (\ref{9-2}) under BCs~(\ref{8}%
) which correspond to the excited states of the condensate. We will
seek each solution in the form of an \textquotedblleft elementary $j_{0}$%
-series\textquotedblright ~\cite{gp1}
\begin{equation}
\Phi _{j_{0}}(x)=\sum\limits_{j=j_{0},3j_{0},5j_{0},\ldots }b_{j}\sqrt{2/L}%
\cdot \sin (k_{j}x),  \quad k_{j}=\pi j/L.  \label{10}
\end{equation}%
The case  $j_{0}=1$ corresponds to the ground state and was
considered in~\cite{gp1}. Below we will find solutions for
$j_{0}=2,3,\ldots ,\infty $ by numerically solving Eqs.~(\ref{9-1}),
(\ref{9-2}) and (\ref{8})  using the Newton method (for a
description of this method, see work~\cite{gp1}).
Since it is impossible to numerically find an infinite number of solutions ($%
j_{0}=2,3,\ldots ,\infty $), we will limit ourselves to the cases
$j_{0}=2, N$, which are of particular interest, and a few random
$j_{0}$. According to our analysis, if the interaction is switched off ($%
\gamma =0$), the solutions transform into the solutions for a free particle
in a box: $\Phi _{j_{0}}(x)|_{\gamma \rightarrow 0}\rightarrow b_{j_{0}}%
\sqrt{2/L}\,\sin (k_{j_{0}}x)$. To clarify the physical meaning of
the solutions $\Phi _{j_{0}}(x)$ for $\gamma >0$, we will compare
them with the exact solutions obtained by the Bethe ansatz.

In the GP and GP$_{N}$ approaches, the condensate energy for the
state $\Phi _{j_{0}}(x)$ is determined by the formula
\begin{equation}
E_{\mathrm{GP}}=\int\limits_{0}^{L}dx\left\{ -\frac{\hbar ^{2}}{2m}\Phi
_{j_{0}}^{\ast }(x)\frac{\partial ^{2}}{\partial x^{2}}\Phi
_{j_{0}}(x)+qc|\Phi _{j_{0}}(x)|^{4}\right\} ,  \label{11}
\end{equation}%
where $q=1$ for the GP approach and $q=1-1/N$ for the GP$_{N}$ one~\cite{gp1}%
. In the exact Bethe-ansatz approach, the energy of the system is
\begin{equation}
E_{\mathrm{Bethe}}=k_{1}^{2}+k_{2}^{2}+\ldots +k_{N}^{2},  \label{12}
\end{equation}%
where $|k_{j}|$ are solutions of the system of Gaudin's equations \cite%
{gaudin1971,gaudinm}
\begin{equation}
L|k_{p}|=\pi n_{p}+\sum\limits_{j=1}^{N}\left( \arctan {\frac{c}{%
|k_{p}|-|k_{j}|}}+\arctan {\frac{c}{|k_{p}|+|k_{j}|}}\right) |_{j\neq
p},\quad p=1,\ldots ,N.  \label{13}
\end{equation}%
Here the quantum numbers $n_{p}$ must be natural, i.e.
$n_{p}=1,2,3,\ldots ,\infty $ for each $p=1,2,\ldots ,N$.

\begin{figure}[ht]
\centerline{\includegraphics[width=85mm]{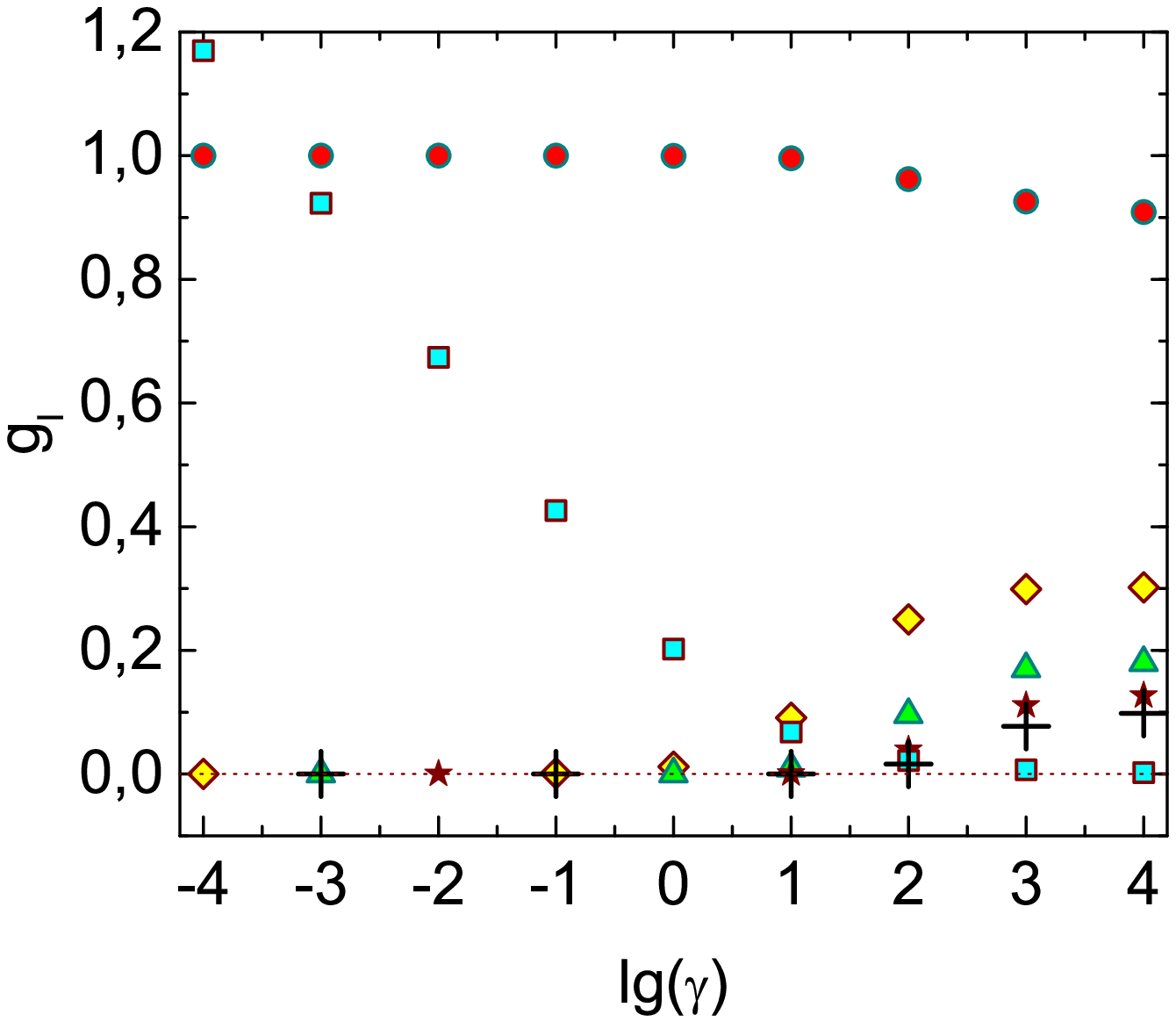} } \caption{
[Color online] Coefficients $g_{1}$ (circles),
$g_{2}$ (diamonds), $g_{3}$ (triangles), $g_{4}$ (stars), $g_{5}$ (crosses), and $%
g_{l_{m}}\equiv \tilde{\epsilon}=\epsilon /[2\bar{\rho}c(1-N^{-1})]$
(squares) obtained within the GP approach for the first stationary
excited state of the condensate $\Phi _{j_{0}=2}(x)$ for $N=2$,
$\bar{\rho}=1$, and various $\gamma $'s. Instead of $g_{l_{m}} $,
the values of $\lg (g_{l_{m}})/4$ are given. The values of all
$g_{l}$ obtained for $\gamma =10^{4}$ practically
coincide with the seed ones $g_{l<l_{m}}=\frac{2\sqrt{2}}{\pi (2l-1)}$, $%
g_{l_{m}}=1 $. The dotted line marks the zero level, $g_{l}=0$. In
the GP approach, exactly the same solutions $g_{l}(\gamma )$ are obtained for $%
j_{0}=3,4,\ldots $ if $N=j_{0}$ (in this case, $g_{l}$ depend on $\gamma =c/%
\bar{\rho}$, but do not on $c$ and $\bar{\rho}$ separately; see
equations (51) and (52) in~\cite{gp1}).
 \label{fig1exgN2}}
\end{figure}

\section{Main hypothesis}

It is impossible to compare the solution $\Phi
_{j_{0}}(x)$~(\ref{10}) with all exact Bethe-ansatz solutions
because their number is infinite. However, we may try to intuitively
guess which exact solution corresponds to the function $\Phi
_{j_{0}}(x)$.

We proceed from the fact that each solution of Eq.~(\ref{9-1}) or (\ref{9-2}%
) describes a state where all $N$ atoms are in the condensate. We know that $%
\Phi _{1}(x)$ describes the ground state and corresponds to the
exact Bethe-ansatz solution for~\cite{gp1} $n_{1}=n_{2}=\ldots
=n_{N}=j_{0}=1$ (or $n_{p\leq N}=j_{0}=1$ for short). The states
$\Phi _{j\geq 2}(x)$
describe the excited states of the condensate. Knowing the complete $N$%
-particle wave function, which is a solution of the exact $N$-particle Schr%
\"{o}dinger equation, one can represent each excited state of the
system as a set of interacting elementary
excitations~\cite{fey1954,holes2020}. Each $\Phi _{j_{0}\geq 2}(x)$
describes a state with the maximum possible number (i.e. $N$) of
atoms in the condensate  and therefore must correspond to a peculiar
ensemble of elementary excitations that forms such a condensate of
atoms. \textit{It is natural to assume that $\Phi _{j_{0}\geq 2}(x)$
corresponds to a state with the maximum possible number of identical
elementary excitations. In this case, it is as if the excitations
focus the atoms in a single-particle state that corresponds to the
structure of these excitations.} In sections 6 and 7 we will show
that $\Phi _{j_{0}\geq 2}(x)$ corresponds to the exact Bethe-ansatz
solution with $n_{p\leq N}=j_{0}$, which describes a condensate of
$N$ identical elementary excitations, each of which has the
quasimomentum $\pi (j_{0}-1)/L$.

\begin{figure}
\includegraphics[width=0.47\textwidth]{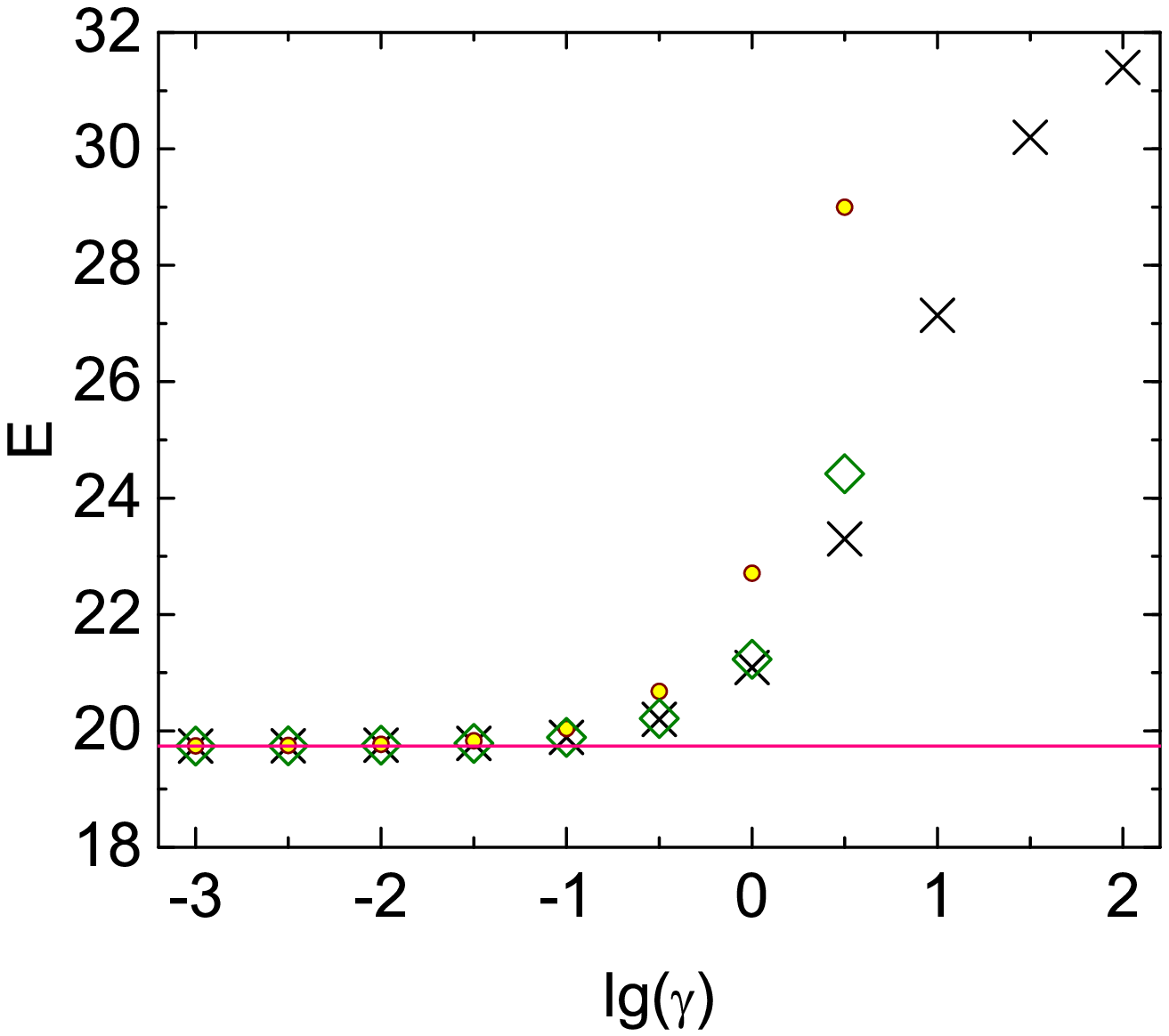}
\hfill
\includegraphics[width=0.47\textwidth]{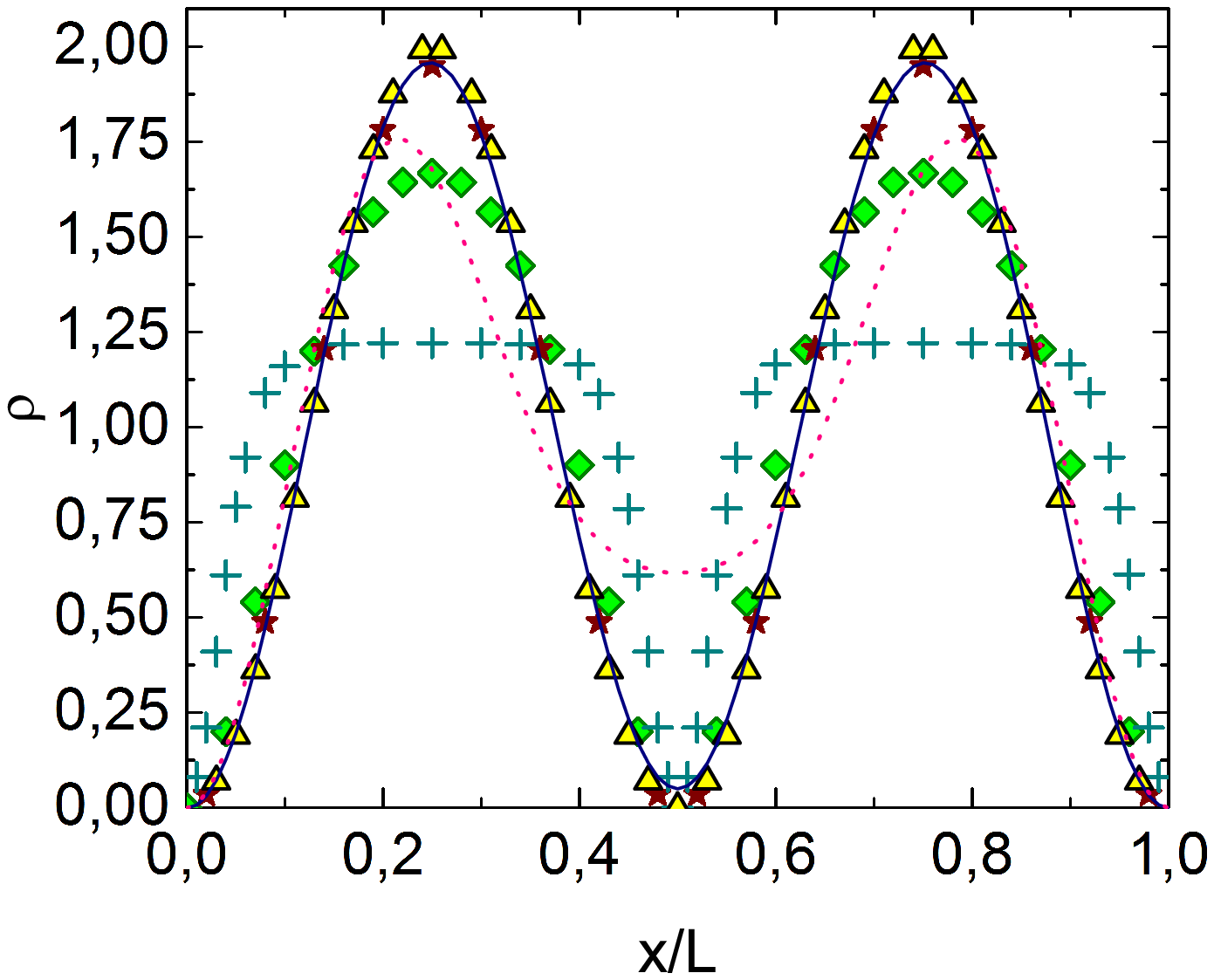}
\\
\parbox[t]{0.47\textwidth}{
\caption{ [Color online]  System energy $E(\gamma )$
calculated for $N=2$ and $\bar{\rho}=1$ in different approaches: the energies $E_{\mathrm{%
GP}}$ (circles) and $E_{\mathrm{GP_{N}}}$ (diamonds) for the state
$\Phi _{j_{0}=2}(x)$, and the exact Bethe-ansatz solution for
$n_{p\leq N}=j_{0}=2$ (crosses). The solid line marks the energy $E=N(j_{0}\pi /L)^{2}$ of $%
N$ free particles for $j_{0}=2$.
 \label{fig2exeN2}} } \hfill
\parbox[t]{0.47\textwidth}{
\caption{ [Color online]  Particle density profile $\rho (x)$ found
for $N=2$ and $\bar{\rho}=1$ in various approaches: the GP solutions
for the state $\Phi _{2}(x)$ at $\gamma =10^{-4}$ (triangles), $1$
(stars), $10$
(diamonds),  $100$ (crosses); the exact Bethe-ansatz solution for $%
n_{1}=n_{2}=j_{0}=2$ and $\gamma =1$ (solid curve),  $\gamma =10$
(dotted curve). The free-particle solution $\rho (x)=2\bar{\rho}\sin
^{2}{(j_{0}\pi x/L)}$ for $j_{0}=2$ and the exact Bethe-ansatz
solution for $\gamma =10^{-4}$ virtually coincide with the curve
marked with triangles and therefore are not shown.
 \label{fig3exroN2}} }
\end{figure}

\section{Two-domain state: $j_{0}=2$}

In this section we study the solution $\Phi _{2}(x)$ which describes
the first excited state of the condensate ($j_{0}=2$). The method of
numerical solution of the GP~(\ref{9-1}) (or
GP$_{N}$~(\ref{9-2})) equation satisfying BCs~(\ref{8}) and the normalization condition $%
\int_{0}^{L}dx|\Phi (x)|^{2}=N$ is described in work~\cite{gp1}. We
will see below that the solution $\Phi _{2}(x)$ at $N=2$ corresponds
to a perfect crystal composed of two atoms, and at $N\gg 1$ it
corresponds to a Lieb's
\textquotedblleft hole-like\textquotedblright\ quasiparticle~\cite{lieb1963}%
. For what follows it is convenient to denote $b_{j_{0}(2l-1)}=\sqrt{N}%
f_{j_{0}(2l-1)}=\sqrt{N}g_{l}$ and write $\Phi
_{j_{0}}(x)$~(\ref{10}) in the form \cite{gp1}
\begin{equation}
\Phi _{j_{0}}(x)=\sqrt{2\bar{\rho}}\sum\limits_{l=1,2,\ldots ,\infty
}g_{l}\sin [\pi j_{0}(2l-1)x/L].  \label{4-1}
\end{equation}%
The function $\Phi _{j_{0}}(x)$ is periodic with the period
$\triangle x=2L/j_{0}$.

\subsection{Solution for $N=2$}

Consider the case $N=j_{0}=2$ and $\bar{\rho}=1$. The solutions for
the coefficients $g_{l}$ from Eq.~(\ref{4-1}) for different $\gamma
$'s are shown in Fig.~\ref{fig1exgN2}. The dependence of $g_{l}$ on
$\gamma $ is similar to that for the ground state (see~\cite{gp1}).
The figure shows the solutions calculated only in the GP approach;
the GP$_{N}$ solutions are similar.

The dependences $E(\gamma )$ are shown in Figs.~\ref{fig2exeN2} and \ref%
{fig10-16exeN}(a); they are similar to  $E(\gamma )$
for the ground state~\cite{gp1}. Note that if $\gamma \ll 1$, the energies $E_{%
\mathrm{GP}}$ and $E_{\mathrm{GP}_{N}}$ [formula (\ref{11}) with $q=1$ and $%
q=1-1/N$, respectively] are close to the exact energy $E_{\mathrm{Bethe}}$ (%
\ref{12}) for $n_{p\leq N}=j_{0}$. In this case,
$E_{\mathrm{GP}_{N}}$ coincides with $E_{\mathrm{Bethe}}$ with high
accuracy.

Figure~\ref{fig3exroN2} demonstrates the particle density profile
$\rho (x)=|\Phi _{j_{0}}(x)|^{2}$ \cite{gp1} for the GP solution
$\Phi _{2}(x)$ at $N=2$, $\bar{\rho}=1$, and different $\gamma $'s.
The GP$_{N}$ solution gives very close profiles; they are not shown
because they would be visually indistinguishable from the GP
profiles. One sees that $\rho (x)$ has a two-domain structure. For
$\gamma \leq 1$, the profile $\rho (x)$ is close to the
free-particle profile $\rho (x)=2\bar{\rho}\sin ^{2}{(j_{0}\pi
x/L)}$ and to the exact $\rho (x)$-solution~\cite{mt2022} found by
the Bethe ansatz. The GP solution $\rho (x)$ differs from the exact
one in that the GP
solution vanishes at $x=0; L/2; L$, whereas the exact solution only at $%
x=0; L$ (the both properties hold for all examined values of $\gamma
$: $10^{-4}\leq \gamma \leq 100$).

\begin{figure}[ht]
\centerline{\includegraphics[width=85mm]{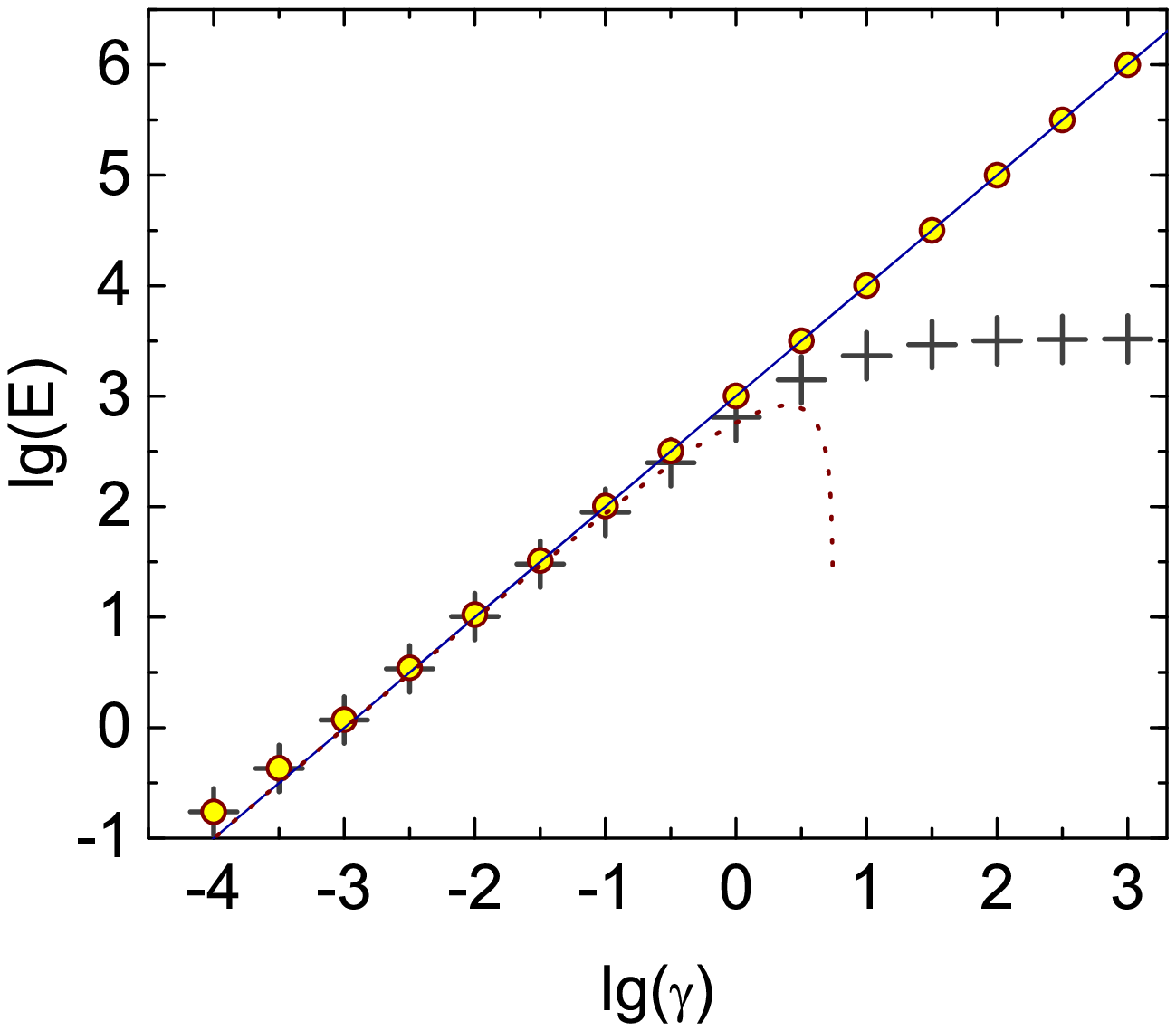} }
\caption{ [Color online] System energy $E(\gamma )$
calculated for $N=10^{3}$ and $\bar{\rho}=1$ by different methods: $E_{%
\mathrm{GP}_{N}}$ for the state $\Phi _{2}(x)$ (circles), the exact energy $%
E_{\mathrm{Bethe}}$ for $n_{j\leq N}=j_{0}=2$ (crosses), and the
Bogoliubov ground-state energy  $E_{0}=N\bar{\rho}^{2}\gamma \left( 1-\frac{4%
\sqrt{\gamma }}{3\pi }\right) $ (dotted curve). The solid line shows
the dependence $E=N\bar{\rho}^{2}\gamma $.
 \label{fig5exeN1000}}
\end{figure}

\subsection{Solution for $N=1000$}

The $E_{\mathrm{GP}_{N}}(\gamma )$ curves for the state $\Phi _{2}(x)$ for $%
N=1000$ are shown in Fig.~\ref{fig5exeN1000}. The GP approach gives
close solutions that would be visually indistinguishable from the
GP$_{N}$ ones.
One can see that at $\gamma \lesssim 0.1$, the energy $E_{\mathrm{GP}%
_{N}} $ is close to the exact energy $E_{\mathrm{Bethe}}$ for
$n_{j\leq
N}=j_{0}=2$. It is interesting that at $\gamma \lesssim 0.1$, the energy $E_{%
\mathrm{GP}_{N}}$ in Fig.~\ref{fig5exeN1000}~is close to the
Bogoliubov
energy of the ground state. This is probably due to the fact that at $%
\gamma \lesssim 0.1$ the function $\Phi _{2}(x)$ changes smoothly (see Fig.~\ref%
{fig6exroN1000}), so that the kinetic energy is low and the total
energy of the system is mostly potential. At $\gamma \lesssim 0.1$
the latter is close to the Bogoliubov energy $E_{0}$, which is also
mostly potential.

\begin{figure}[ht]
\centerline{\includegraphics[width=85mm]{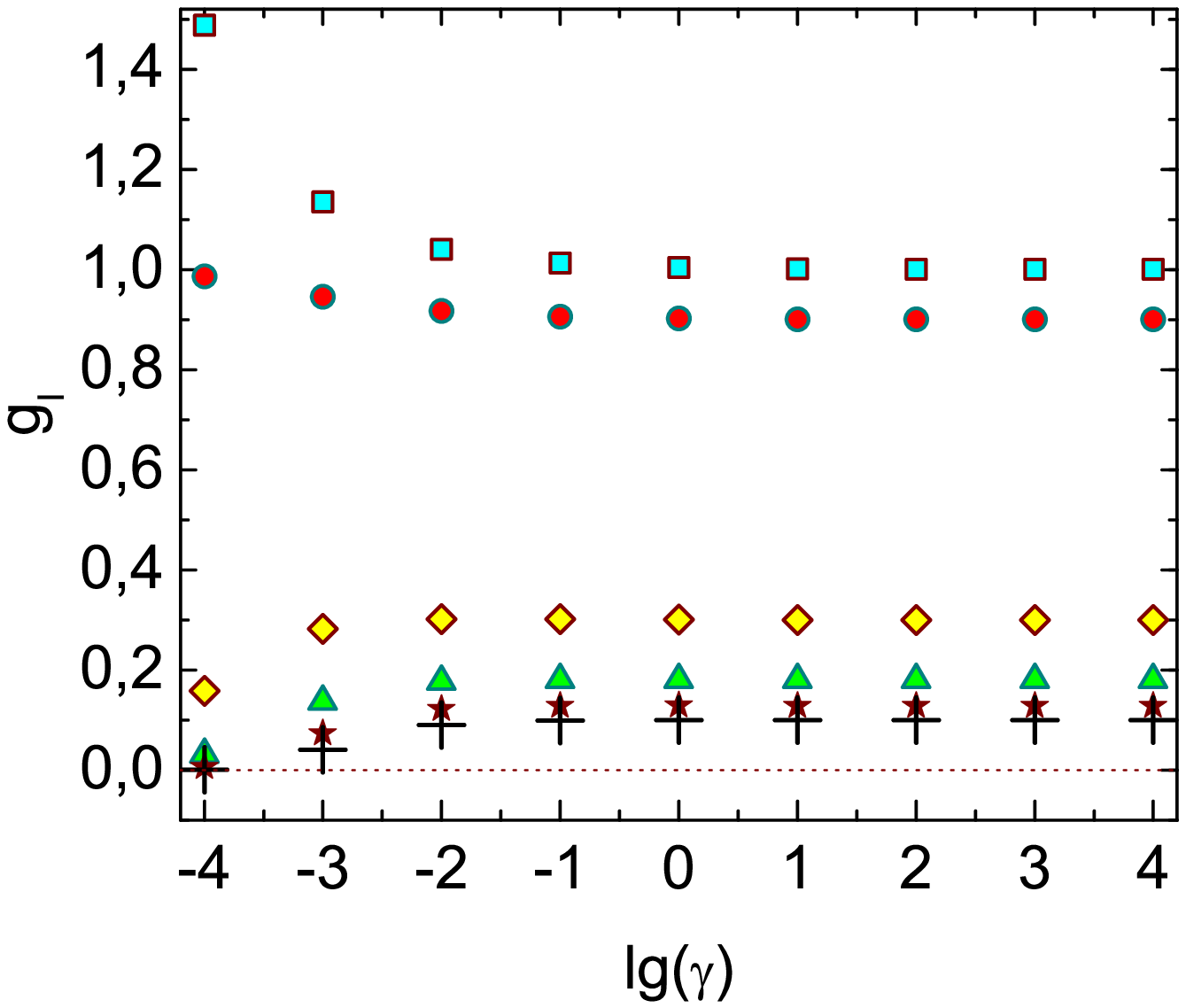} }
\caption{ [Color online] Coefficients $g_{1}$ (circles), $%
g_{2}$ (diamonds), $g_{3}$ (triangles), $g_{4}$ (stars), $g_{5}$
(crosses), and $g_{l_{m}}\equiv \tilde{\epsilon}$ (squares) obtained
within the GP approach
for the state $\Phi _{2}(x)$ at $N=1000$, $\bar{\rho}=1$, and different $%
\gamma $'s. In contrast to Fig.~\ref{fig1exgN2}, the values of $g_{l_{m}}$ (instead of $%
\lg (g_{l_{m}})/4$) are shown. When $\lg (\gamma )\gtrsim 0$ the
values of
the coefficients $g_{l}$ practically coincide with $g_{l<l_{m}}=\frac{2\sqrt{%
2}}{\pi (2l-1)}$, $g_{l_{m}}=1$. The dotted line marks the zero level $%
g_{l}=0$.
 \label{fig4exgN1000}}
\end{figure}

The values of the coefficients $g_{l}$ for various $\gamma $'s are
shown in Fig.~\ref{fig4exgN1000}. The $g_{l}(\gamma )$ curves for
all $l$ are close to their ground-state counterparts for $N=1000$,
see~\cite{gp1}. The figure shows only the solutions obtained in the
GP approach, because the GP$_{N}$ solutions are very close.

\begin{figure}[ht]
\centerline{\includegraphics[width=85mm]{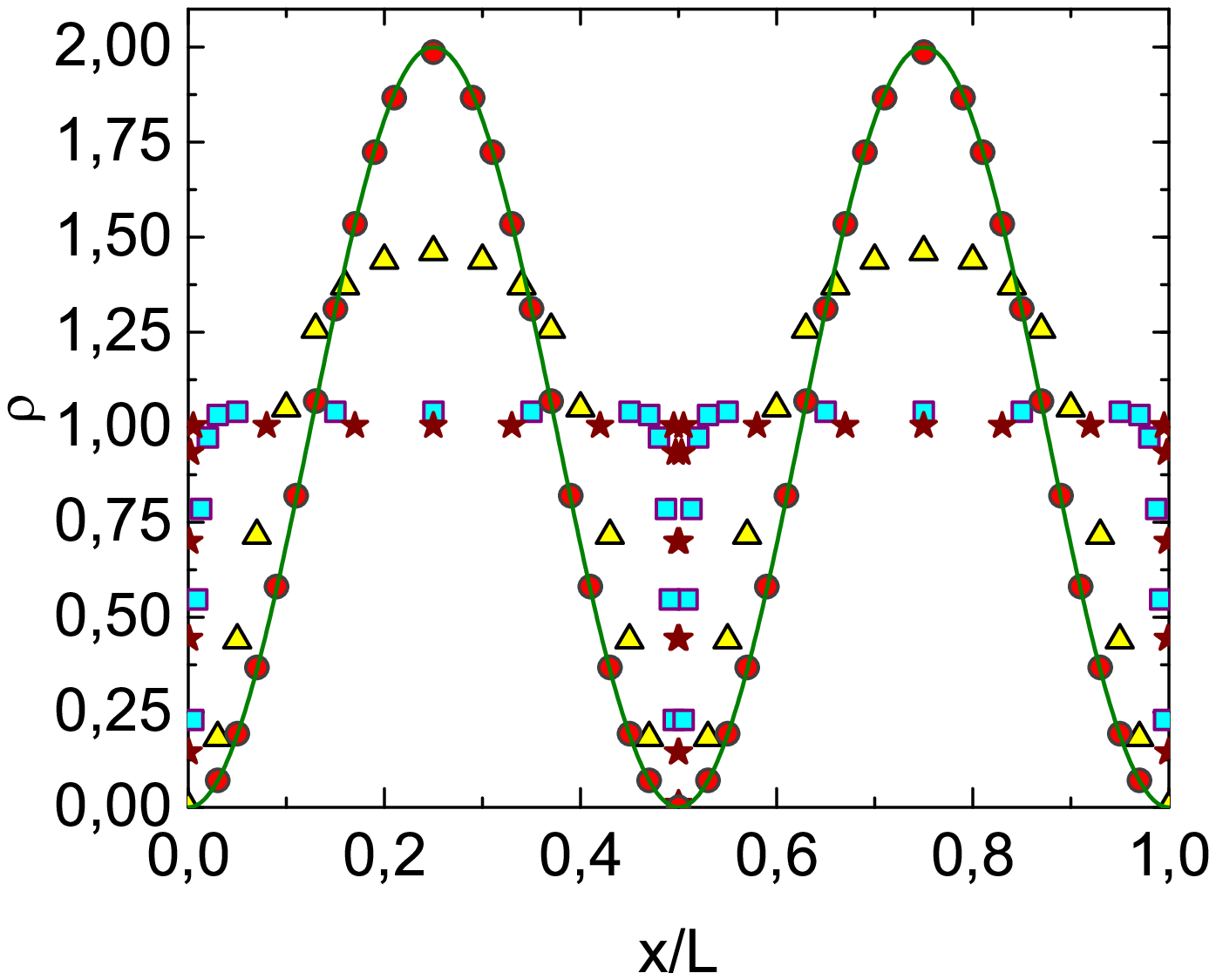} }
\caption{ [Color online] Particle density profiles $%
\rho(x)=|\Phi_{2}(x)|^{2}$ obtained in the GP approach for $N=1000$, $\bar{%
\rho} =1$, and $\gamma=10^{-6}$ (circles), $10^{-4}$ (triangles), $%
10^{-2}$ (squares),  $ 1$ (stars). The solid line marks free
particle
curve $\rho(x)=2\bar{\rho}\sin^{2}{(j_{0}\pi x/L)}$ for $j_{0}=2$, $\bar{%
\rho}=1$.
 \label{fig6exroN1000}}
\end{figure}

Figure~\ref{fig6exroN1000} shows the particle density profiles $\rho
(x)=|\Phi _{j_{0}}(x)|^{2}$ for the GP solution $\Phi _{2}(x)$ at
$N=1000$ and various
$\gamma $'s. The GP$_{N}$ solution gives a very close $\rho (x)$. As in Fig.~%
\ref{fig3exroN2}, the $\rho (x)$ curve has a two-domain structure, and for $%
\gamma N^{2}\lesssim 1$ (the regime of near-free particles
\cite{gp1}) it is close to the free particle curve $\rho (x)=2\bar{\rho}\sin ^{2}{(j_{0}\pi x/L)%
}$.

Figures~\ref{fig2exeN2}, \ref{fig10-16exeN}(a), \ref{fig3exroN2}, and \ref%
{fig5exeN1000} indicate that for $N=2(1000)$ and $\gamma \lesssim
0.1$, the solution $\Phi _{2}(x)$ corresponds to the exact
Bethe-ansatz solution for $N=2(1000)$ and $n_{j\leq N}=2$. In
section~7 we will see that the numbers $n_{j\leq N}=2$ correspond to
the Lieb's \textquotedblleft
hole\textquotedblright . As can be seen from Figs.~\ref{fig3exroN2} and \ref%
{fig6exroN1000}, these are solutions with two-domain profiles $\rho
(x)$. A Bethe-ansatz solution with a two-domain profile $\rho
_{N}(x_{N})$ was previously obtained in work~\cite{syrwid2015} for a
periodic 1D system of spinless bosons.

\begin{figure}
\includegraphics[width=0.47\textwidth]{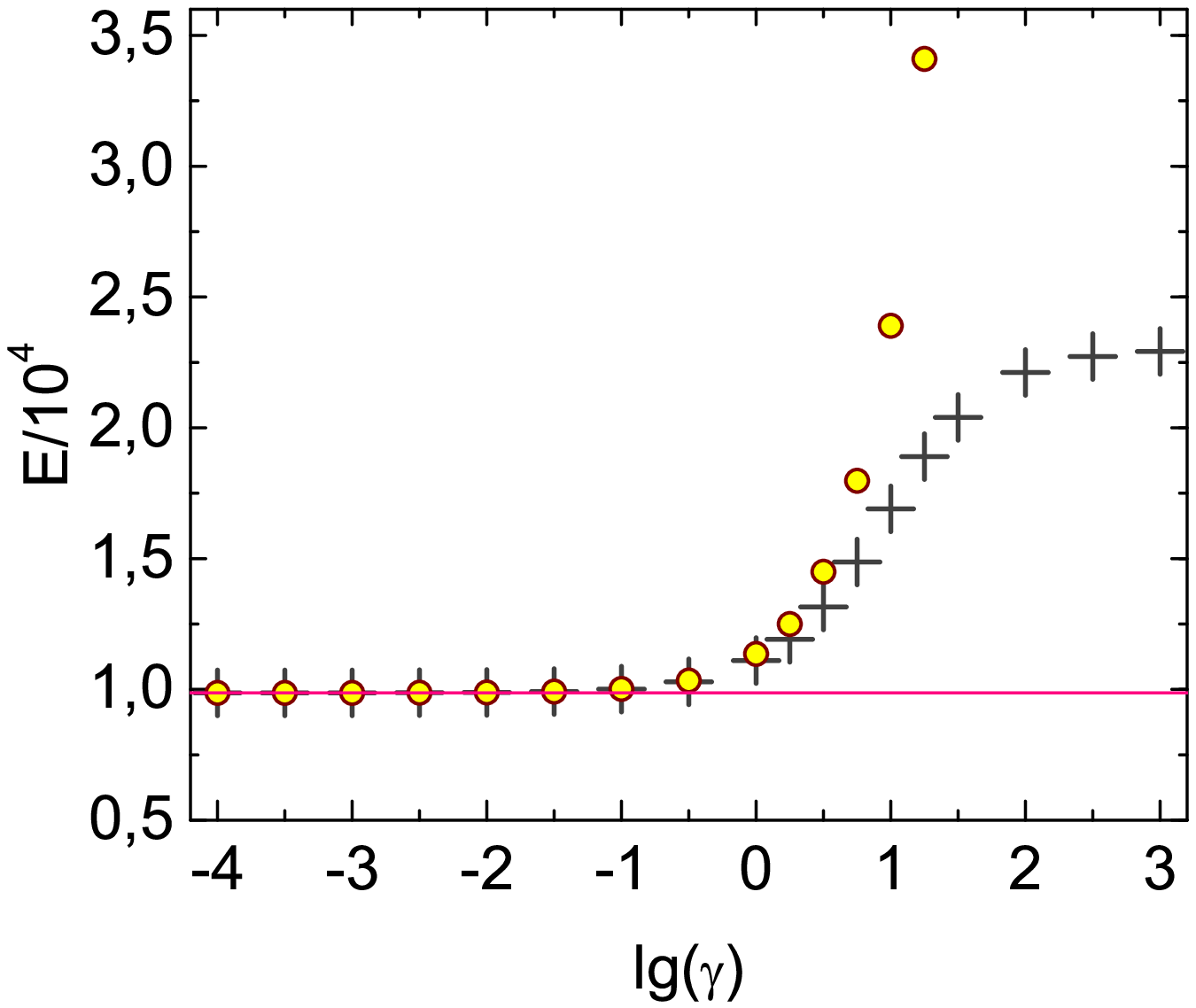}
\hfill
\includegraphics[width=0.47\textwidth]{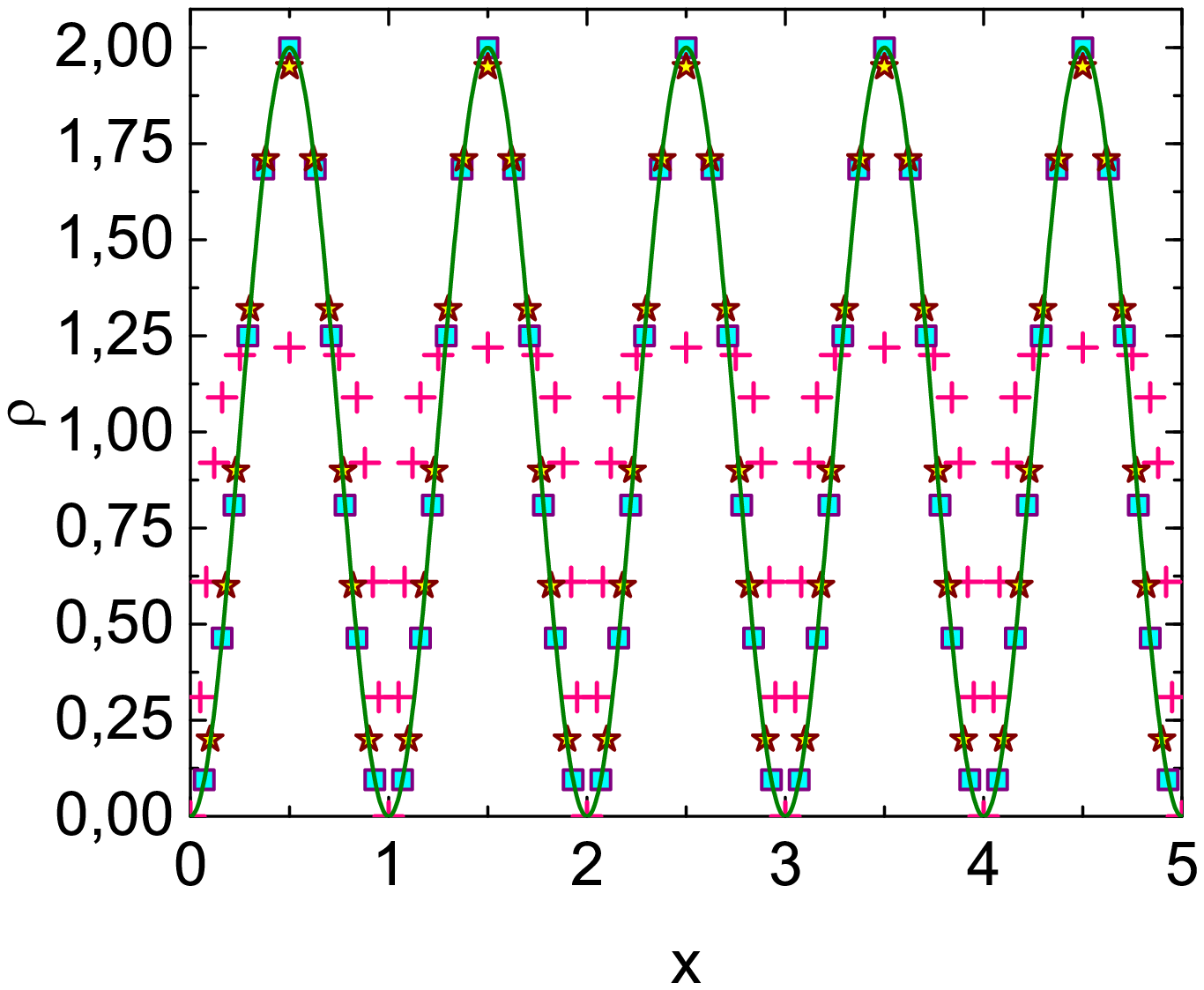}
\\
\parbox[t]{0.47\textwidth}{
\caption{ [Color online]  The energy $E(\gamma )$ of the
crystal-like excited state of the system obtained by different
methods for $N=L=1000$ (the value of $E$ is divided by $10^{4}$):
the energy $E_{\mathrm{GP}_{N}}(\gamma )$ of
the state $\Phi _{j_{0}=N}(x)$ (circles); the exact energy $E_{\mathrm{%
Bethe}}(\gamma )$ for $n_{j\leq N}=j_{0}=N$ (crosses); the solid
line marks the dependence $E(\gamma )=N(j_{0}\pi /L)^{2}$ for the
system of $N$ free particles for $j_{0}=N$.
 \label{fig7excryse}} } \hfill
\parbox[t]{0.47\textwidth}{
\caption{ [Color online]  Particle density profiles $\rho (x)=|\Phi
_{1000}(x)|^{2}$ for the crystal-like excited state of the
condensate calculated in the GP$_{N}$ approach for $j_{0}=N=L=1000$ and $%
\gamma =10^{-2}$ (squares), $1$ (stars), and $10^{2}$ (crosses). The
solid line marks the free-particle curve $\rho (x)=2\bar{\rho}\sin
^{2}{(j_{0}\pi x/L)}$ for $j_{0}=1000$, $\bar{\rho}=1$; with a high
accuracy, it coincides with the curve marked by squares.
 \label{fig8excrysro}} }
\end{figure}

\section{Solution for a perfect crystal: $j_{0}=N$}

Figures~\ref{fig7excryse} and \ref{fig8excrysro} show solutions for
the excited state $\Phi _{j_{0}}(x)$ of the condensate  for
$j_{0}=N=1000$. In the GP approach, the coefficients $g_{l}(\gamma
)$ for $j_{0}=N=1000$ coincide with their counterparts for
$j_{0}=N=2$ (see Fig.~\ref{fig1exgN2}) because of the scaling
properties of the system of equations for $g_{l}$~\cite{gp1}. In the
GP$_{N}$ approach, those equations contain an additional factor
$1-1/N$, which violates the scaling.

It is clear from Fig.~\ref{fig7excryse} that if $\gamma \lesssim
0.1$, the
energy $E_{\mathrm{GP}_{N}}$ is close to the exact energy $E_{\mathrm{Bethe}%
} $ obtained for the quantum numbers $n_{j\leq N}=j_{0}=N$. The
energy $E_{\mathrm{GP}}$ is very close to $E_{\mathrm{GP}_{N}}$.

The particle density profiles $\rho (x)$ calculated for $N=L=1000$ and different $%
\gamma $'s (see Fig.~\ref{fig8excrysro}) show that this solution
corresponds to \textit{a perfect crystal} with one atom per a
lattice site. The system
occupies the interval $x\in \lbrack 0,L]=[0,1000]$, and Fig.~\ref%
{fig8excrysro} demonstrates the first five sites (domains). All $N$
domains are identical, i.e. they are a repetition in space of the
first domain (with $\rho (x)=0$ at the points $x=0,1,2,\ldots
,L-1,L$). Such properties result from the fact that when $x$ is
replaced by $x+L/j_{0}$, the function $\Phi _{j_{0}}(x)$~(\ref{4-1})
does not change its magnitude but changes its sign. Therefore, $\rho
(x)=|\Phi _{j_{0}}(x)|^{2}$ is periodic with the period $\triangle
x=L/j_{0}=1$.

Thus, the solution $\Phi _{2}(x)$ for $N=2$ corresponds to a perfect
crystal consisting of two atoms, and $\Phi _{1000}(x)$ for $N=1000$
corresponds to a perfect crystal of 1000~atoms. The solution $\Phi
_{2}(x)$ is close
to the exact one for $\gamma \lesssim 1$ in the GP$_{N}$ approach, and for $%
\gamma \lesssim 0.1$ in the GP one, whereas the solution $\Phi
_{1000}(x)$ is close to the exact one for $\gamma \lesssim 0.1$ in
both approaches.

Some approximate crystalline solutions with a condensate of atoms
have already been found in the literature \cite%
{gross1958,gross1960,luca,coniglio1969,kirz,nep,shlyapa2015,andreev2017,mtfragm2,fil2020}%
. The exact crystalline solution for a 1D few-boson system with
point interaction was obtained in~\cite{mt2022}, and it corresponds
to the solution $\Phi _{j_{0}=N}(x)$ above. Its energy is always
higher than the energy of the ground state (the latter corresponds
to a liquid); therefore, such a crystal is not a supersolid. If the
interatomic potential is non-point,  the crystalline solution may
correspond to the ground state of the system.

\begin{figure}
\begin{center}
\includegraphics[width=.47\textwidth]{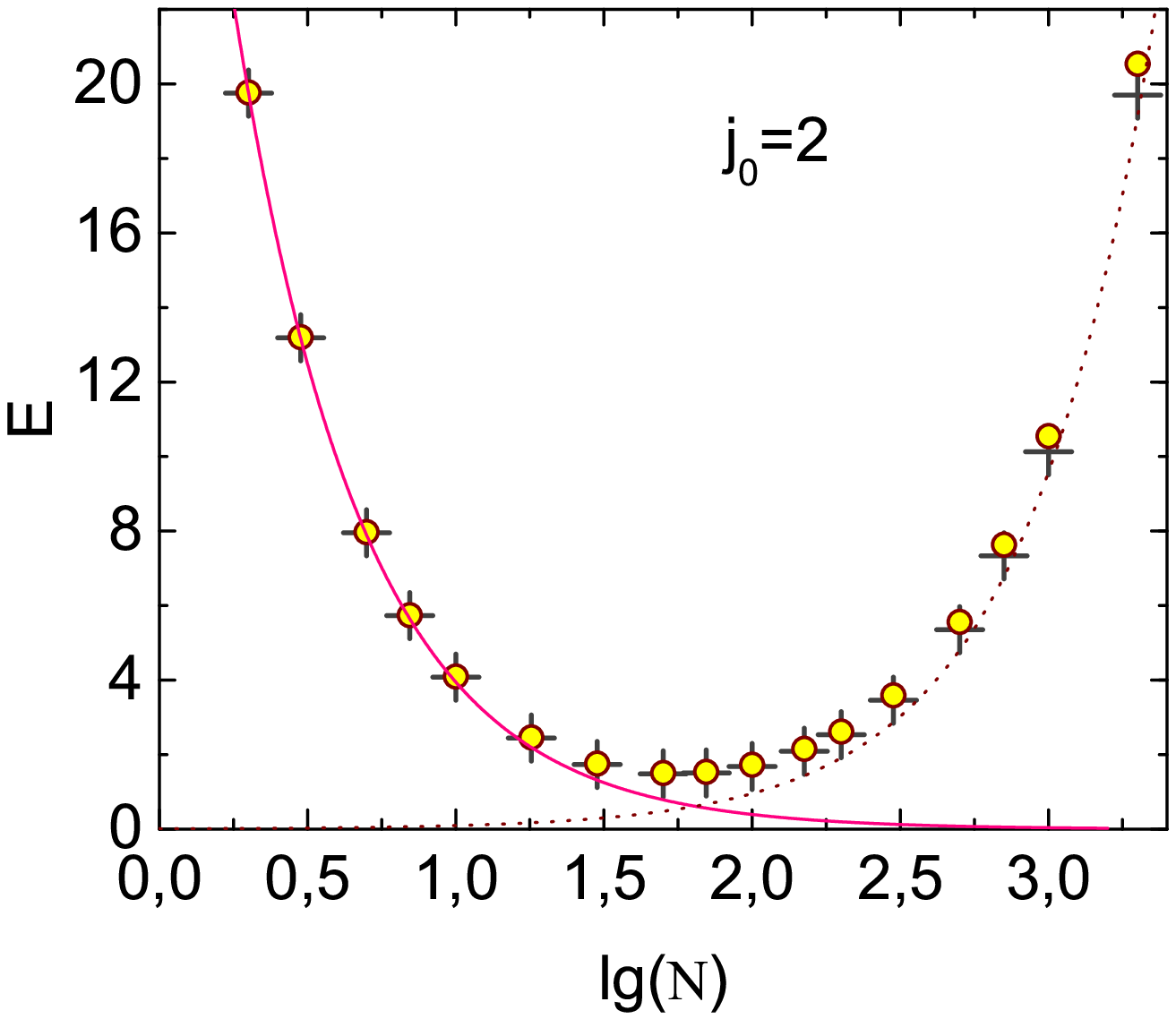}
\hspace{.04\textwidth}
\includegraphics[width=.47\textwidth]{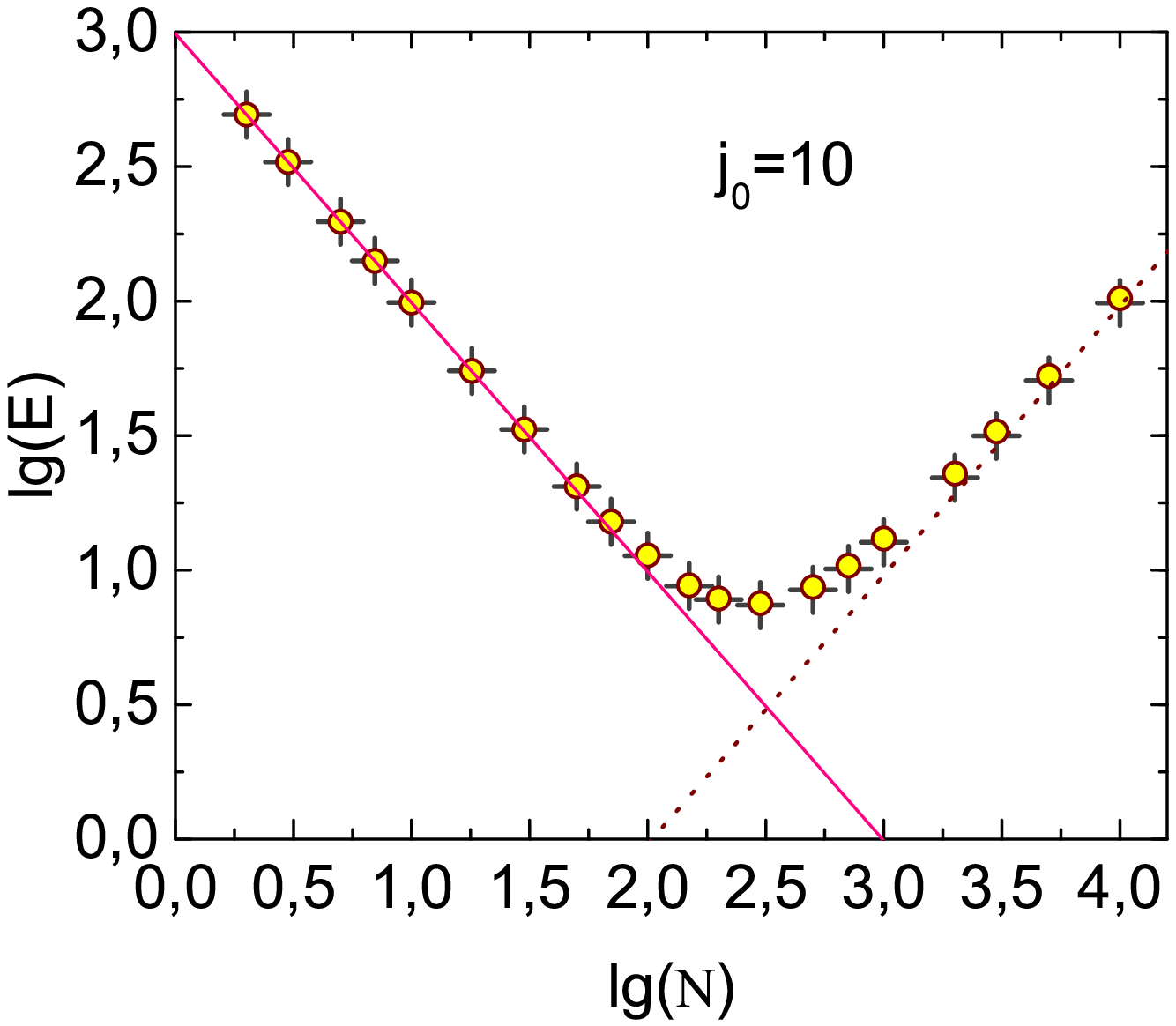}\\
\includegraphics[width=.47\textwidth]{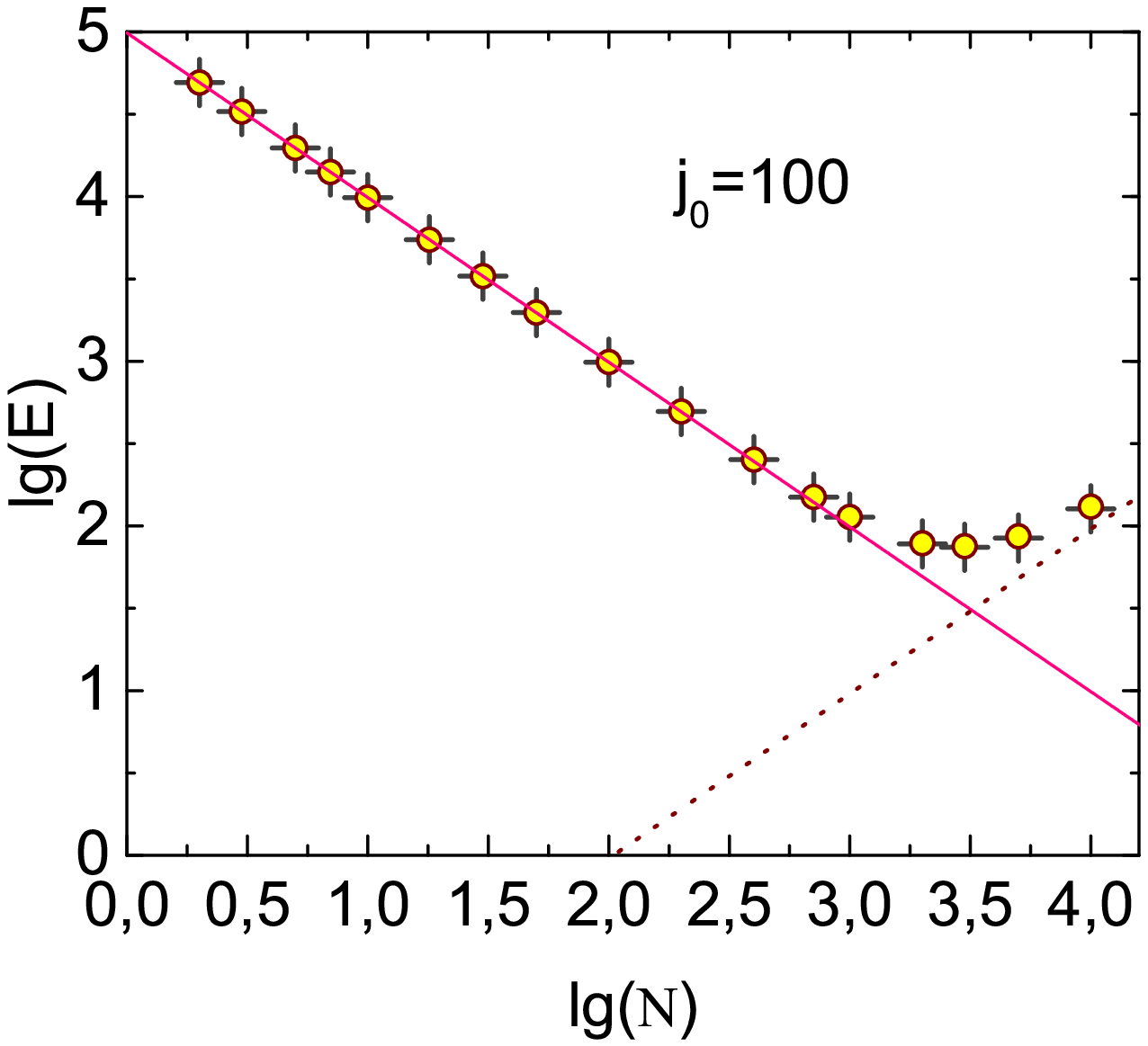}
\hspace{.04\textwidth}
\includegraphics[width=.47\textwidth]{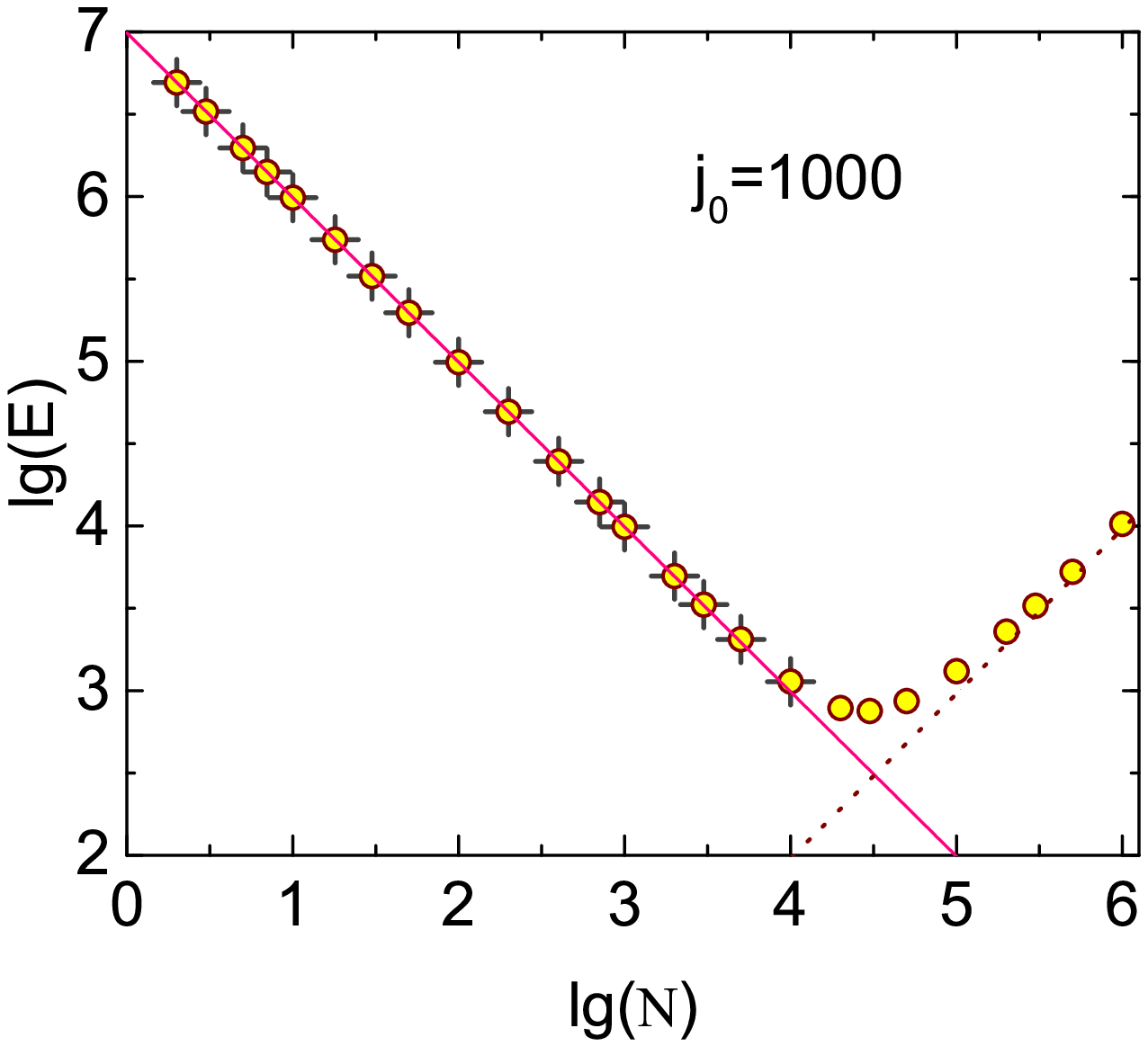}
\caption{ [Color online] The dependence $E(N)$ for $c=0.01$, $%
\bar{\rho}=1$, and $j_{0}=2,10,100,1000$, calculated with different
methods: the energy $E_{\mathrm{GP}}(N)$ of the condensate state
$\Phi _{j_{0}}(x)$ (circles), the exact energy
$E_{\mathrm{Bethe}}(N)$ for the
state with $n_{j\leq N}=j_{0}$ (crosses), the Bogoliubov ground-state energy $E_{0}(N)=N%
\bar{\rho}^{2}\gamma \left( 1-\frac{4\sqrt{\gamma }}{3\pi }\right) $
(dotted curves), and the energy of $N$ free particles $%
E(N)=N(j_{0}\pi /L)^{2}$ (solid curves). In the panel corresponding to $%
j_{0}=1000$, the solution $E_{\mathrm{Bethe}}$ is shown only for
$N\leq 10^{4}$ (we were unable to solve the Gaudin's equations
(\ref{13}) for $N>10^{4}$ due to the large computing time).}
\label{fig12-15exen}
\end{center}
\end{figure}

\section{Correspondence between the solutions of the Gross-Pitaevskii
equation and the exact Bethe-ansatz solutions}

In work \cite{gp1} it was shown that the solution $\Phi _{1}(x)$ of
the GP (GP$_{N}$) equation, corresponding to the ground state of
condensate, for $\gamma \lesssim 0.1$ coincides with the exact
Bethe-ansatz solution with the quantum numbers $n_{j\leq N}=j_{0}=1$
(the latter solution
corresponds to the ground state of the system for any $\gamma \geq 0$~\cite%
{gaudin1971,mt2015,mtjpa2017}). Which quantum numbers $(n_{1},\ldots
,n_{N})$ in the Bethe-ansatz approach correspond to the solutions
$\Phi_{j_{0}>1}(x)$? The dependences $E(\gamma )$ and $E(N)$ (see Figs.~\ref%
{fig2exeN2}, \ref{fig5exeN1000}, \ref{fig7excryse}, and
\ref{fig12-15exen}) show that if $N<10$, the solution $\Phi
_{j_{0}>1}(x)$ should correspond to the exact solution with
$n_{j\leq N}=j_{0}$. For $N\gtrsim 10$, the energy $E$ of the system
determined in the GP and GP$_{N}$ approaches has a rather large
error. As a result, the solution $\Phi _{j_{0}>1}(x)$
can be assigned both to the state with $n_{j\leq N}=j_{0}$ and to other $%
\{n_{j}\}$-states with similar energies.

Fig.~\ref{fig12-15exen} shows some interesting properties. First,
when $N\lesssim 10j_{0}$, the energy $E_{\mathrm{GP}}$ is very close
both to
the exact energy $E_{\mathrm{Bethe}}$ and the energy of $N$ free particles $%
E=N(j_{0}\pi /L)^{2}$ [in this Section, $E_{\mathrm{Bethe}}$ does
not denote the Bethe-ansatz energy of any ($n_{1},n_{2},\ldots
,n_{N}$)-state, but only of the state with $n_{j\leq N}=j_{0}$,
where $j_{0}$ is the same as in $\Phi _{j_{0}}(x)$]. Those two
properties are apparently  related to the fact that at $c=0.01$,
$\bar{\rho}=1$, and $N\lesssim 10j_{0}$ the regime of near-free
particles (see~\cite{gp1}) is realized. One can also see from
Fig.~\ref{fig12-15exen} that when $N\gtrsim 100j_{0}$, the energy
$E_{\mathrm{GP}}$ is close to the Bogoliubov ground-state energy,
although $E_{\mathrm{GP}}$ corresponds to a highly excited state of
the system. This strange property can be explained as follows. The
solution $\Phi _{j_{0}}(x)$ corresponds to $j_{0}$ domains. The
inequality $N\gtrsim 100j_{0}$ means that each domain contains more
than 100~atoms. A numerical analysis shows that in this case the
main contribution to $E_{\mathrm{GP}}$ is given by the potential
energy of interatomic interaction (i.e. by uniform intradomain
regions rather than interdomain ones with a high kinetic energy),
which is similar to the Bogoliubov ground state. We found such
properties to be valid for all examined values of $j_{0}$ and $N$.

\begin{figure}
\begin{center}
\includegraphics[width=.32\textwidth]{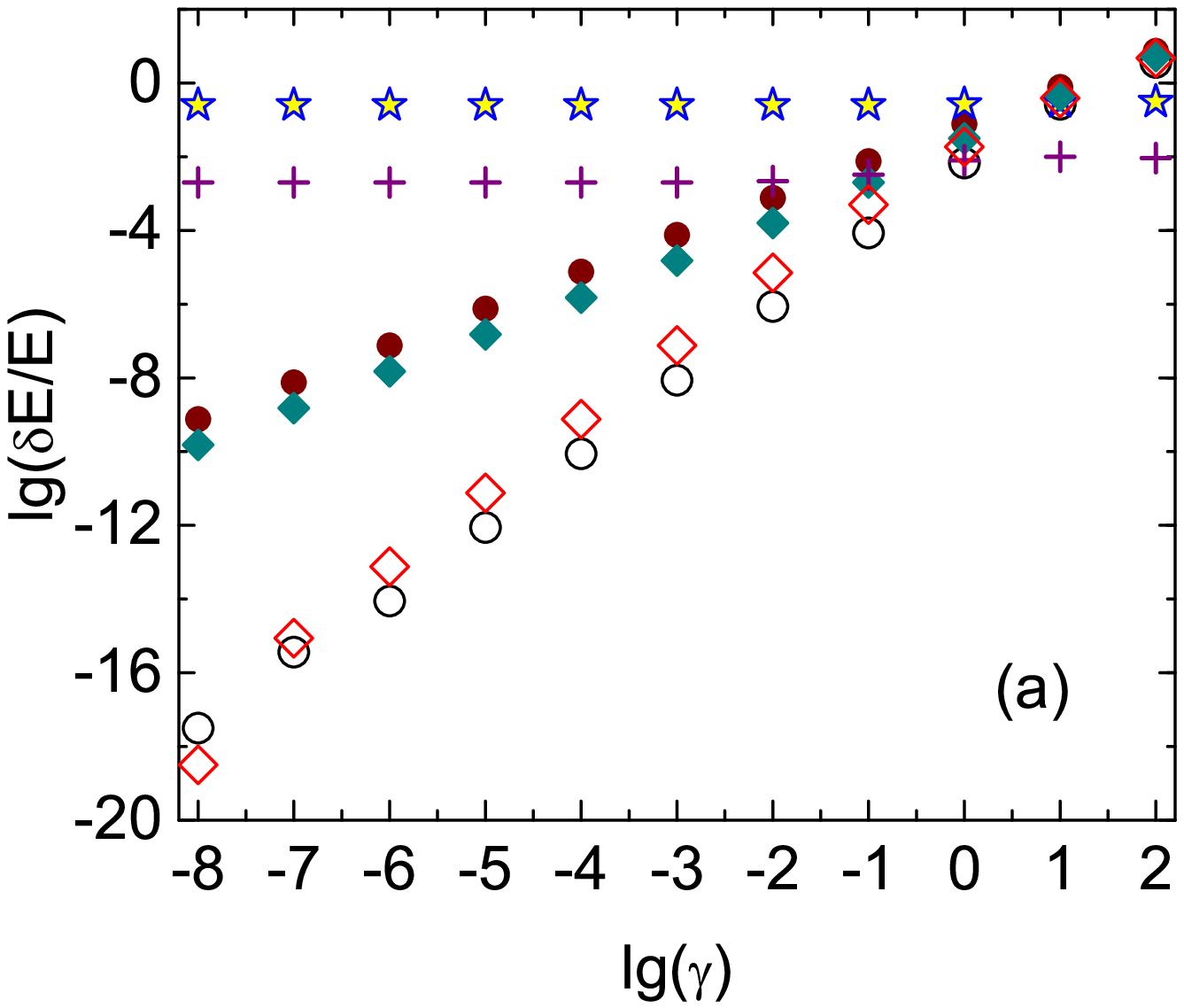}
\includegraphics[width=.32\textwidth]{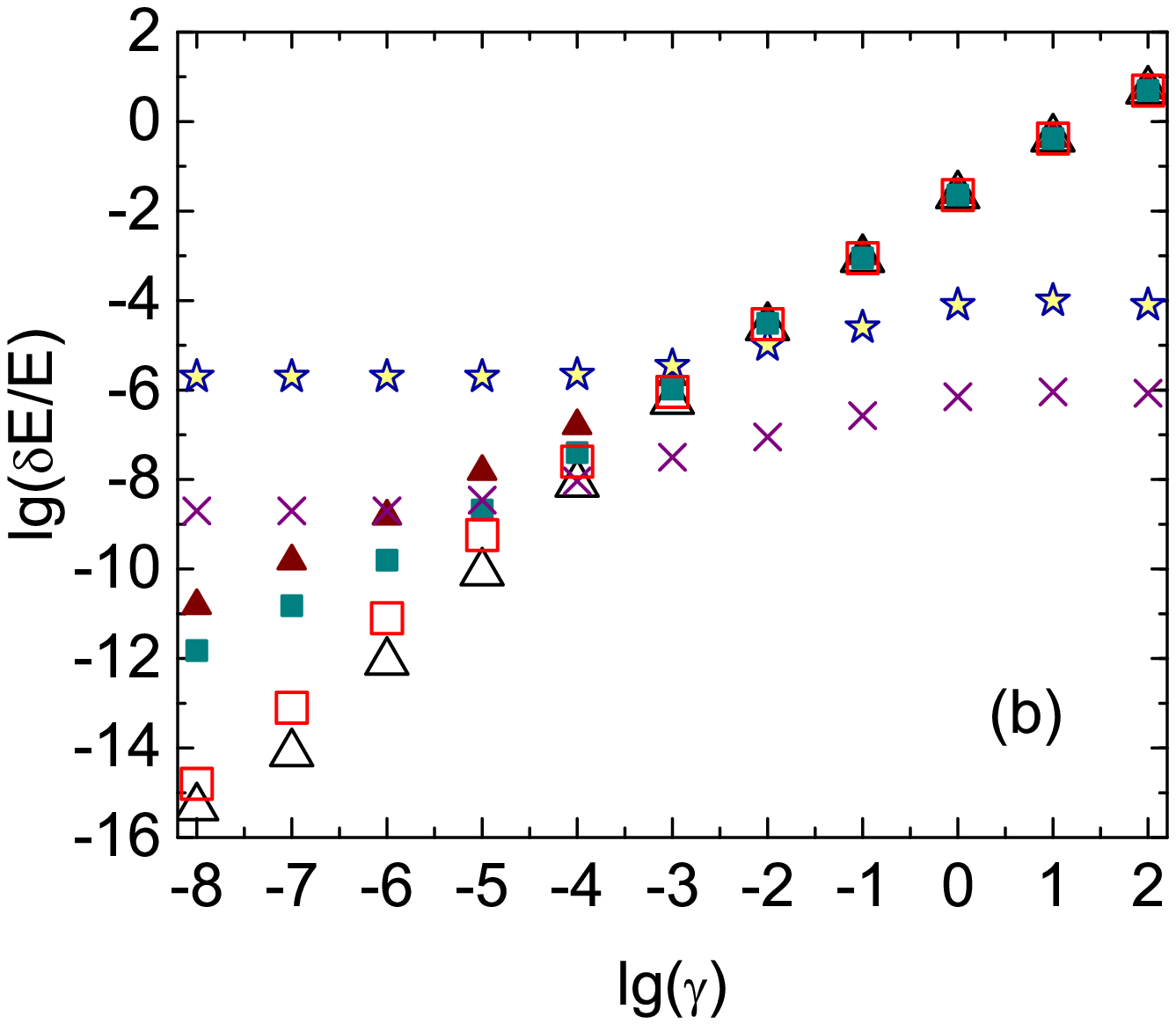}
\includegraphics[width=.32\textwidth]{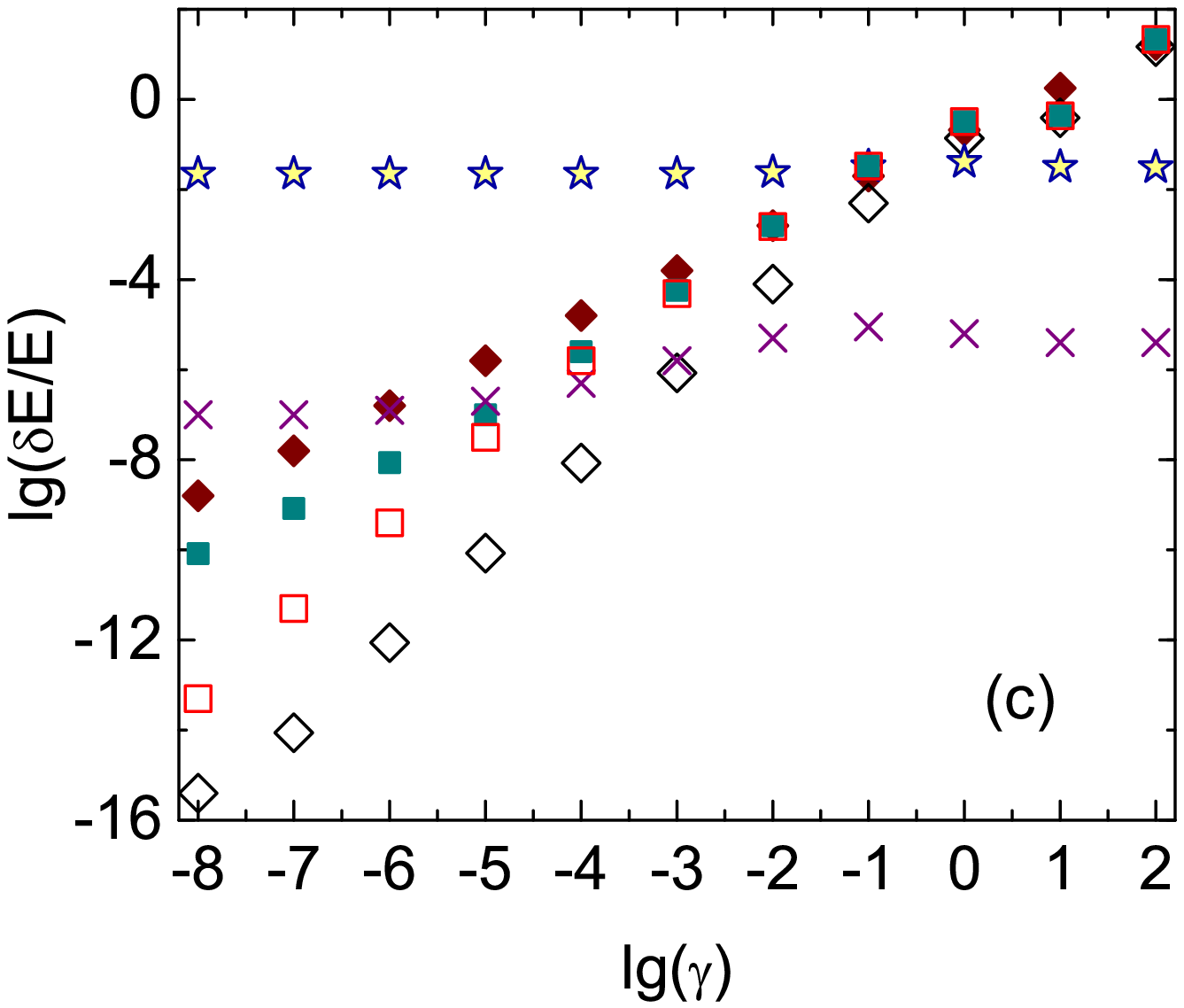}
\caption{ [Color online] Errors of the GP and GP$_{N}$ methods for
the excited states $\Phi _{j_{0}>1}(x)$ at $\bar{\rho}=1$.
\textbf{(a)}~$\gamma $-Dependences of $\lg {\frac{E_{\mathrm{GP}%
}-E_{\mathrm{Bethe}}}{E_{\mathrm{Bethe}}}}$ for $N=10$ (solid
diamonds) and $N=2$
(solid circles), $\lg {\frac{|E_{\mathrm{GP}_{N}}-E_{\mathrm{Bethe}}|}{E_{%
\mathrm{Bethe}}}}$ for $N=10$ (hollow diamonds ) and  $N=2$ (hollow
circles),
and $\lg {\frac{E_{\mathrm{Bethe}}^{+}-E_{\mathrm{Bethe}}}{E_{\mathrm{Bethe}}%
}}$ for $N=2$ (stars) and $10$ (crosses). \textbf{(b)}~$\gamma $-Dependences of $\lg {%
\frac{E_{\mathrm{GP}}-E_{\mathrm{Bethe}}}{E_{\mathrm{Bethe}}}}$ for
$N=100$
(solid triangles) and $1000$ (solid squares), $\lg {\frac{|E_{\mathrm{GP}%
_{N}}-E_{\mathrm{Bethe}}|}{E_{\mathrm{Bethe}}}}$ for $N=100$ (hollow
triangles ) and $1000$ (hollow squares), and $\lg {\frac{E_{\mathrm{Bethe}%
}^{+}-E_{\mathrm{Bethe}}}{E_{\mathrm{Bethe}}}}$ for $N=100$ (asterisks) and $%
1000$ (crosses). In panels~(a) and (b), each of $E_{\mathrm{GP}}$, $E_{%
\mathrm{GP}_{N}}$, $E_{\mathrm{Bethe}}$, and
$E_{\mathrm{Bethe}}^{+}$
corresponds to $j_{0}=N$. \textbf{(c)}~$\gamma $-Dependences of $\lg {\frac{|E_{%
\mathrm{GP}_{N}}-E_{\mathrm{Bethe}}|}{E_{\mathrm{Bethe}}}}$, $\lg {\frac{E_{%
\mathrm{GP}}-E_{\mathrm{Bethe}}}{E_{\mathrm{Bethe}}}}${,} and $\lg {\frac{E_{%
\mathrm{Bethe}}^{+}-E_{\mathrm{Bethe}}}{E_{\mathrm{Bethe}}}}$ for $N=10$ and $%
j_{0}=3$ (hollow diamonds, solid diamonds, and stars, respectively), and for $%
N=1000$ and $j_{0}=135$ (hollow squares, solid squares, and crosses,
respectively). The quantum numbers $n_{p}$ for the exact solutions $E_{%
\mathrm{Bethe}}$ and $E_{\mathrm{Bethe}}^{+}$ are given in the text
(see section~6). } \label{fig10-16exeN}
\end{center}
\end{figure}

Since $\Phi _{1}(x)$ is reliably identified as the ground state, the
quantity $\frac{|E_{\mathrm{GP}_{N}}-E_{\mathrm{Bethe}}|}{E_{\mathrm{Bethe}}}
$ for $\Phi _{1}(x)$~\cite{gp1} virtually determines the accuracy of the GP$%
_{N}$ approach. Figure~\ref{fig10-16exeN} shows that the values of
this quantity for all studied excited states are \textit{smaller}
than its value for the ground state (see Fig.~7 in~\cite{gp1})
provided the same $N=2$, $10$, $100$, $1000$ and the same $\gamma
\lesssim 0.1$. This fact indicates that in the case $\gamma \lesssim
0.1$, the solution $\Phi _{j_{0}>1}(x)$ coincides with the
Bethe-ansatz solution for $n_{j\leq N}=j_{0}$ within the error
limits of the GP$_{N}$ approach.

In addition, Fig.~\ref{fig10-16exeN} shows the values of  $E_{%
\mathrm{Bethe}}^{+}-E_{\mathrm{Bethe}}$  for different $\gamma $'s.
Here $E_{\mathrm{Bethe}}^{+}$ is the energy of the state
($n_{1}=j_{0}-1, n_{2}=n_{3}=\ldots =n_{N-1}=j_{0}, n_{N}=j_{0}+1$)
[if $N=2$, then
$n_{1}=j_{0}-1, n_{N}=j_{0}+1$]. A numerical analysis shows that for any $N\geq 2$ and $%
j_{0}>1$, the level $E_{\mathrm{Bethe}}^{+}$ is one of the closest
(by energy) to the level $E_{\mathrm{Bethe}}$. For each $N\geq 2$,
there are Bethe-ansatz states whose energy $E$ is closer to the
level $E_{\mathrm{Bethe}}$ at a
given $\gamma \lesssim 1$ (the number of such states is of the order of $%
Nj_{0}$; in this case, $\left\vert E-E_{\mathrm{Bethe}}\right\vert \sim |E_{%
\mathrm{Bethe}}^{+}-E_{\mathrm{Bethe}}|$ as a rule, although $\left\vert
E-E_{\mathrm{Bethe}}\right\vert \ll |E_{\mathrm{Bethe}}^{+}-E_{\mathrm{Bethe}%
}|$ for some levels). It is clear that if $\left\vert E_{\mathrm{GP}_{N}}-E_{%
\mathrm{Bethe}}\right\vert \ll |E-E_{\mathrm{Bethe}}|$ for
\textit{all} Bethe-ansatz levels $E$ that are close to
$E_{\mathrm{Bethe}}$, then the GP$_{N}$ approach describes a state
which \textit{exactly coincides} with the state
$n_{j\leq N}=j_{0}$ of the Bethe-ansatz approach. For $N\leq 5$ and $%
\gamma \lesssim 1$, we have numerically verified that $\left\vert E_{\mathrm{%
GP}_{N}}-E_{\mathrm{Bethe}}\right\vert \ll |E-E_{\mathrm{Bethe}}|$
indeed for all Bethe-ansatz levels $E$ and all $j_{0}=2, 3, \ldots,
1000$. When $N>5$, the computing time was too long, so we could not
verify all required Bethe-ansatz levels.

\begin{figure}
\begin{center}
\includegraphics[width=.32\textwidth]{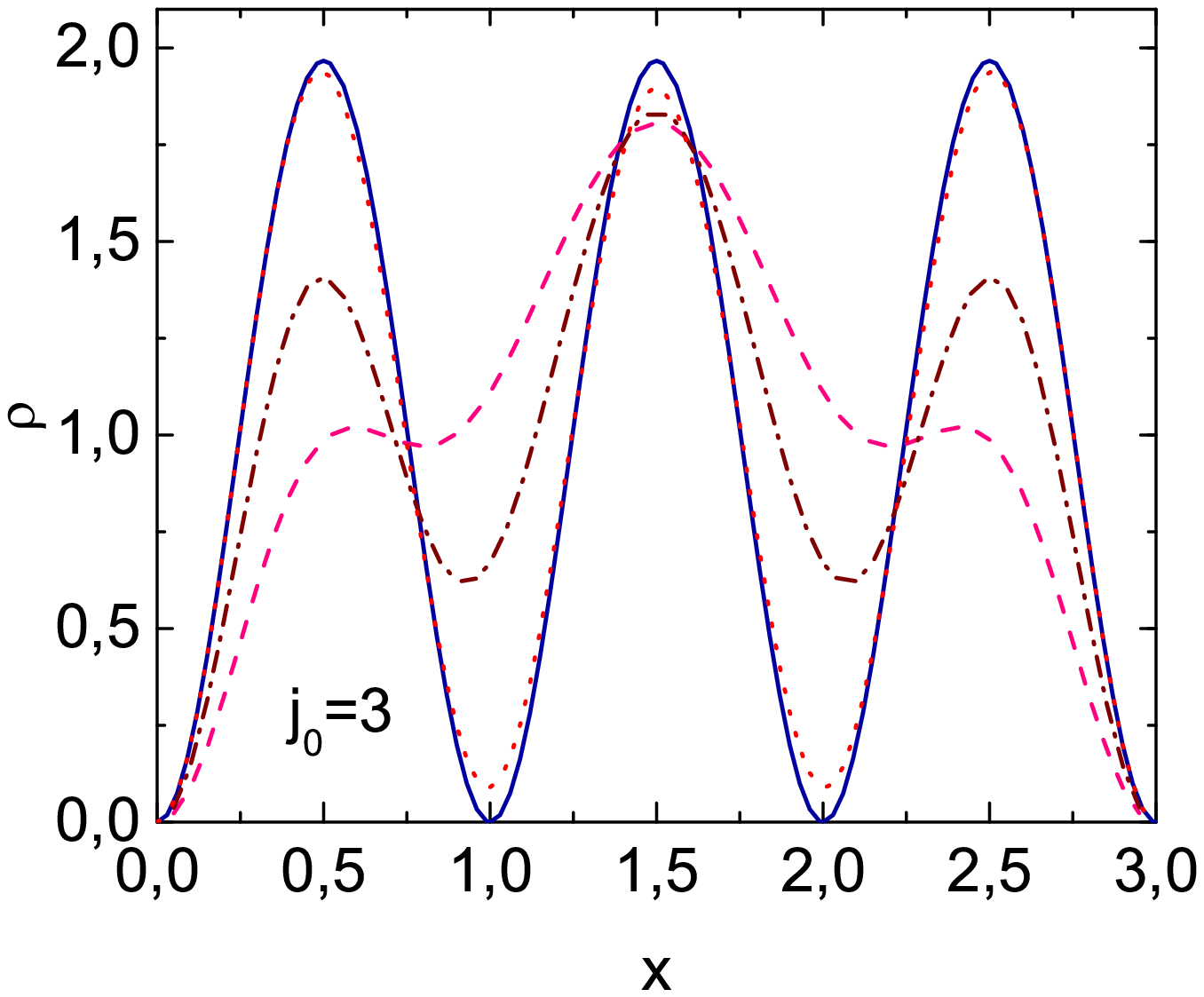}
\includegraphics[width=.32\textwidth]{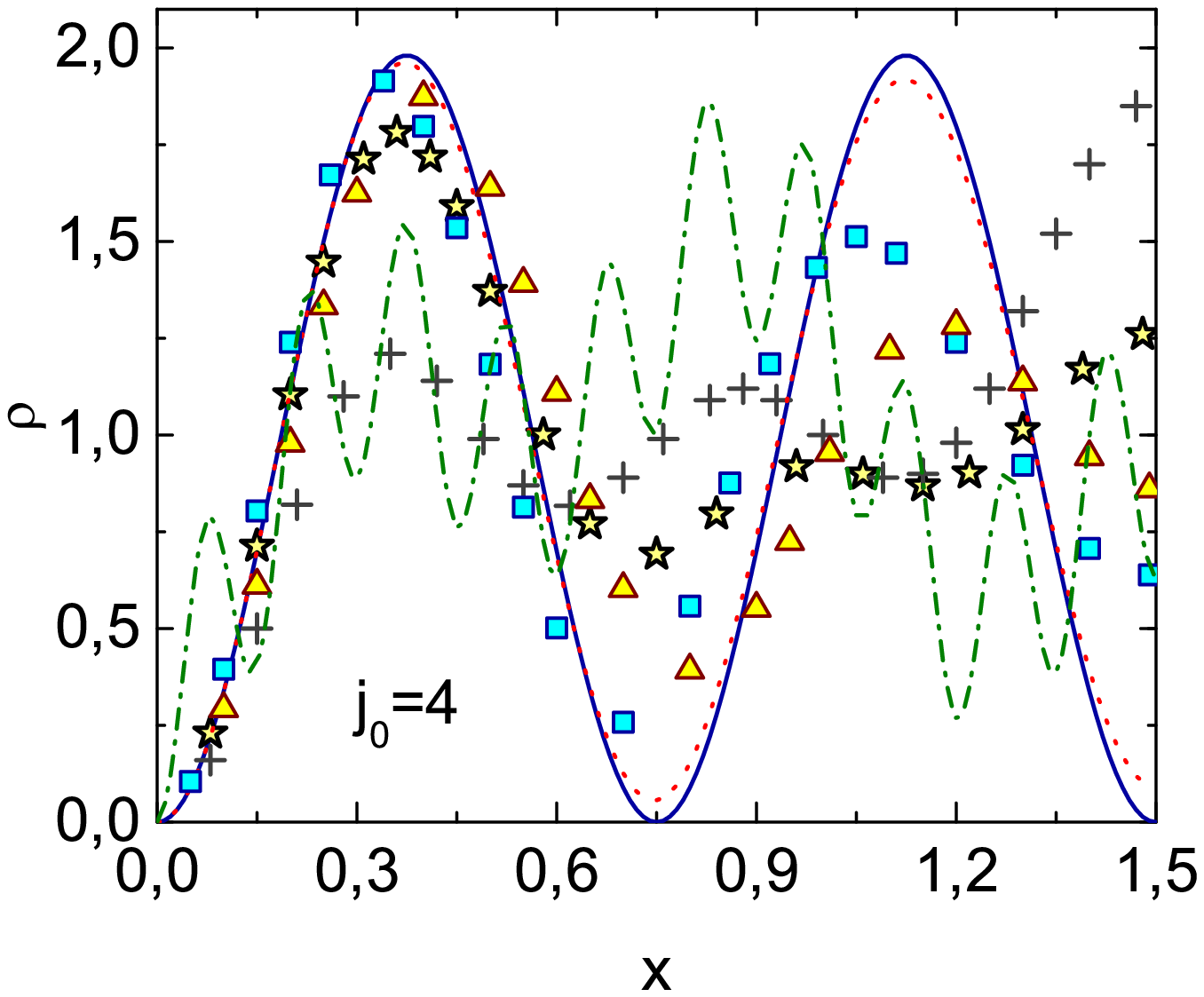}
\includegraphics[width=.32\textwidth]{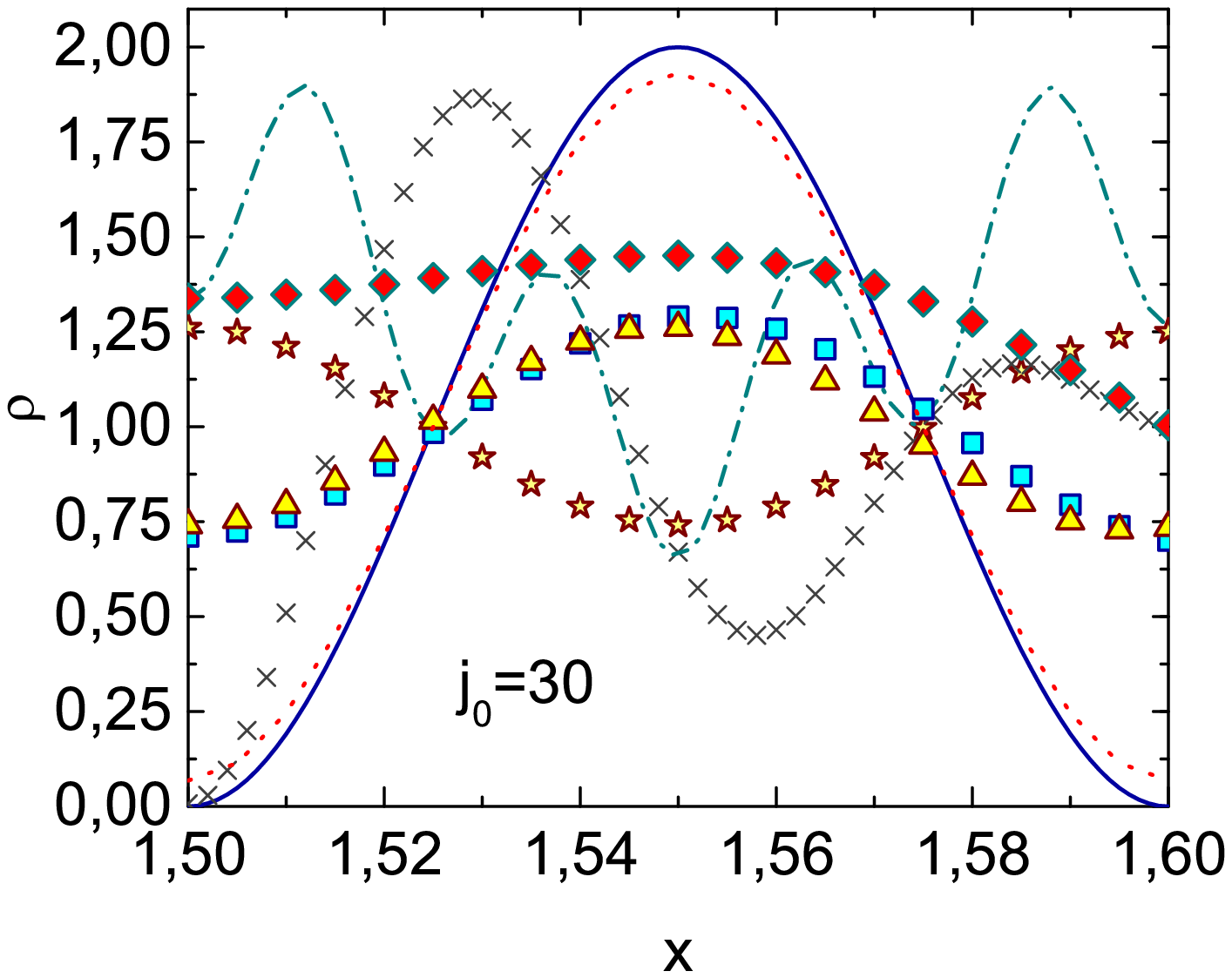}
\caption{ [Color online] Particle density profiles $\rho (x)$
calculated for $N=L=3$, $\gamma =1$, and $j_{0}=3$, $4$, $30$ in
different approaches: the GP$_{N}$ solution for the state $\Phi
_{j_{0}}(x)$ (solid
curves), the exact Bethe-ansatz solutions for the states $%
n_{1}=n_{2}=n_{3}=j_{0}$ (dotted curves); $(n_{1}=j_{0}-1,
n_{2}=j_{0}, n_{3}=j_{0})$ [triangles]; $(n_{1}=j_{0}, n_{2}=j_{0},
n_{3}=j_{0}+1)$ [squares]; and $(n_{1}=j_{0}-1, n_{2}=j_{0},
n_{3}=j_{0}+1)$ [stars]. The $\left( j_{0}=3\right) $-panel also shows the   $%
\rho (x)$-profiles for the Bethe-ansatz states $(n_{1}=1, n_{2}=1,
n_{3}=3)$ [dashed curve] and $(n_{1}=1, n_{2}=3, n_{3}=3)$
[dash-dotted curve]. The $\left( j_{0}=4\right) $-panel additionally
shows the $\rho (x)$-profiles for the states  $(n_{1}=1, n_{2}=3,
n_{3}=5)$ [crosses] and $(n_{1}=2, n_{2}=5, n_{3}=20)$ [dash-dotted
curve]; here $\rho (x)$ is shown in the interval $x\in \lbrack
0,L/2]$; the
profile in the interval $x\in [L/2,L]$ can be obtained using the formula $%
\rho (x)=\rho (L-x)$. The $\left( j_{0}=30\right) $-panel also
demonstrates the $\rho (x)$-profile for the states $(n_{1}=40,
n_{2}=50, n_{3}=60)$ [crosses], $(n_{1}=5, n_{2}=15,
n_{3}=20)$ [diamonds], and $(n_{1}=3, n_{2}=31, n_{3}=120)$ [dash-dotted curve]; $%
\rho (x)$ is shown in the interval $x\in [1.5,1.6]$ corresponding to
the 16th domain of the GP$_{N}$ solution (the whole system is
located within the interval $x\in [0,3]$). In all panels, the dotted
curves almost coincide with the solid ones. } \label{fig18exroN3}
\end{center}
\end{figure}

Important additional information is provided by the profile $\rho
(x)$.  When $\gamma\rightarrow 0$, the Bethe-ansatz for any state
$\{n_{j}\}$ can be written as
$\Psi(x_{1},\ldots,x_{N})=(2i)^{N}\sum_{\sigma}\prod_{j=1}^{N}\sin
(k_{\sigma(j)}x_{j})$ with $k_{\sigma(j)}=\pi n_{\sigma(j)}/L$
\cite{syrwid2017}, and $\Phi_{j_{0}}(x)$ is reduced to
$b_{j_{0}}\sqrt{2/L}\,\sin (k_{j_{0}}x)$ (see Figs.~\ref{fig1exgN2},
\ref{fig4exgN1000} above).  This implies that the domain structure
of the profile $\rho (x)$ of the solution $\Phi_{j_{0}}(x)$ for
$\gamma= 0$ is reproduced \textit{only} by the profile $\rho (x)$ of
the exact Bethe-ansatz solution for $n_{j\leq N}=j_{0}$. If
$\gamma>0$ the picture is less obvious. However, our analysis
in this paper (see Fig.~\ref{fig18exroN3}) and in~%
\cite{mt2022} shows that when $\gamma>0$ the conclusion is the same
(in this case the profile $\rho (x)$ of the solution $\Phi
_{j_{0}}(x)$ consists of $j_{0}$ \textit{identical} domains, whereas
the
profile of the Bethe-ansatz solution for $n_{j\leq N}=j_{0}$ consists of $%
j_{0}$ \textit{almost identical} domains, with the difference
between them being negligibly small if $\gamma \lesssim 1$). Even
the nearest
Bethe-ansatz sets $(n_{1}=j_{0}-1,n_{2}=\ldots =n_{N}=j_{0})$ and $%
(n_{1}=\ldots =n_{N-1}=j_{0},n_{N}=j_{0}+1)$ bring about the
profiles $\rho (x)$ that differ considerably from the profile of the
solution $\Phi _{j_{0}}(x) $, and the profiles $\rho (x)$ for other
$\{n_{j}\}$-sets differ still more strongly from the latter; see
Fig.~\ref{fig18exroN3} (for $N=2$ and $3$, this property was proven
in this work and~\cite{mt2022}; however, it is clear that it should
hold for any $N\geq 2$).

For $N=10$ we found such Bethe-ansatz levels $E$ for which $\left\vert E-E_{%
\mathrm{Bethe}}\right\vert
<|E_{\mathrm{Bethe}}^{+}-E_{\mathrm{Bethe}}|$. All these solutions
had the quantum numbers $\{n_{j}\}$ that differed significantly from
the set $n_{j\leq N}=j_{0}$ (this is also true for all similar
$E$-levels at $N\leq 5$). As a result, the $\rho (x)$ profiles of
such solutions should be very different from the GP$_{N}$ profile
consisting of $j_{0}$ identical domains. Hence, those solutions do
not correspond to the GP$_{N}$ solution. Therefore, for $N>5$,
instead of finding all $E$-levels, we found only the level
$E_{\mathrm{Bethe}}^{+}$, which is close to the level
$E_{\mathrm{Bethe}}$ and has quantum numbers close to $n_{j\leq
N}=j_{0}$.

The condition $\left\vert E_{\mathrm{GP}_{N}}-E_{\mathrm{Bethe}}\right\vert
\ll E_{\mathrm{Bethe}}^{+}-E_{\mathrm{Bethe}}$ is satisfied if $\gamma
N^{2}\lesssim 1$, i.e. in the near-free-particle regime (see Fig.~\ref%
{fig10-16exeN}). Taking into account the properties of the profile
$\rho (x)$ that were described above, we arrive at the conclusion
that for $\gamma N^{2}\lesssim 1$ the solution $\Phi _{j_{0}}(x)$ of
the GP$_{N}$ equation corresponds to the exact Bethe-ansatz solution
with $n_{j\leq N}=j_{0}$, for any $N\geq 2$, $j_{0}=1,2,\ldots
,\infty $, and $\gamma \lesssim 0.1$. For
the values of $\gamma $ and $N$ that  violate the relation $%
\gamma N^{2}\lesssim 1$, the inequality $\left\vert E_{\mathrm{GP}_{N}}-E_{%
\mathrm{Bethe}}\right\vert \ll
E_{\mathrm{Bethe}}^{+}-E_{\mathrm{Bethe}}$ becomes invalid. However,
this is apparently a result of the GP$_{N}$ approach error: from the figures in this article and in~\cite%
{gp1}, one  sees that this error is very small if $\gamma
N^{2}\lesssim 1$ but increases rapidly when moving away from this
parameter region. The error of the GP approach exceeds the error of
the GP$_{N}$ approach, but the general conclusions for the GP
approach are about the same.

It is also possible to find the nodal structure of the wave function
$\Psi (x_{1},\ldots ,x_{N})$, i.e. the set of cells into which the
nodes of the wave function divide the space $(x_{1},\ldots ,x_{N})$.
In the GP (GP$_{N}$) approach we have $\Psi (x_{1},\ldots
,x_{N})=\prod_{j=1}^{N}[\Phi _{j_{0}}(x_{j})/\sqrt{N}]$, and the
expression for the Bethe ansatz $\Psi
(x_{1},\ldots ,x_{N})$ can be found in~\cite%
{gp1,syrwid2021,gaudin1971,gaudinm}. In the cases $N=2$ and $N=3$,
the nodal structure of the Bethe ansatz $\Psi (x_{1},\ldots ,x_{N})$
was explored in~\cite{mt2022}. It turned out that the nodal structure of the GP (GP$%
_{N} $) solution $\Psi (x_{1},\ldots ,x_{N})$ for $j_{0}=N$
coincides with the nodal structure of only one Bethe ansatz $\Psi
(x_{1},\ldots ,x_{N})$, namely, the ansatz with $n_{j\leq N}=j_{0}$.
If $N=2$ or $3$, it is easy to verify that this behavior holds for
$j_{0}\neq N$ as well. It is clear that it should hold for all
$N\geq 2$ and $j_{0}\geq 1$.

Thus, the analysis of the energy levels shows that every solution
$\Phi _{j_{0}}(x)$ corresponds exactly to the Bethe-ansatz solution
for $n_{j\leq N}=j_{0}$ and the same $N$ when $N\leq 5$. If $N>5$,
our analysis of the energy levels only allows us to state that each
solution $\Phi _{j_{0}}(x)$ corresponds to a set of Bethe-ansatz
solutions: the solution for $n_{j\leq N}=j_{0}$ and other
$(n_{1},\ldots ,n_{N})$-solutions with very close energies. The
nodal structure of the functions $\Psi (x_{1},\ldots ,x_{N})$ and
the properties of the profiles $\rho (x)$, which were explored for
$N=2$ and $3$, show that each solution $\Phi _{j_{0}}(x)$
corresponds exactly to the Bethe-ansatz solution for $n_{j\leq
N}=j_{0}$ and the same $N$. There is no doubt that in the case $N>3$
the properties of $\rho (x)$ and the nodal structures are also
similar (although we have not proven this statement). Consequently,
every solution $\Phi _{j_{0}}(x)$ corresponds exactly to the
Bethe-ansatz solution for $n_{j\leq N}=j_{0}$ and the same $N$, for
arbitrary $N\geq 2$ and $j_{0}\geq 1$.

\section{Elementary excitations}

Let us find out which set of elementary collective excitations
(elementary quasiparticles) corresponds to the state $n_{j\leq
N}=j_{0}>1$.
According to L.~Landau's idea, a system of many interacting particles at $%
T\rightarrow 0$ can be considered as a small number of free
quasiparticles~\cite{landau1941}. In a Bose system, several
quasiparticles can be considered as a single one, so quasiparticles
are introduced ambiguously: an infinite number of different
quasiparticle ensembles can be associated with the same system. As
an elementary excitation, one usually calls an indivisible
quasiparticle, i.e. a quasiparticle that cannot be represented as
several interacting quasiparticles. Intuitively, one may expect that
the simplest formulae are obtained by describing a system in the
language of elementary excitations.

\begin{figure}[ht]
\centerline{\includegraphics[width=85mm]{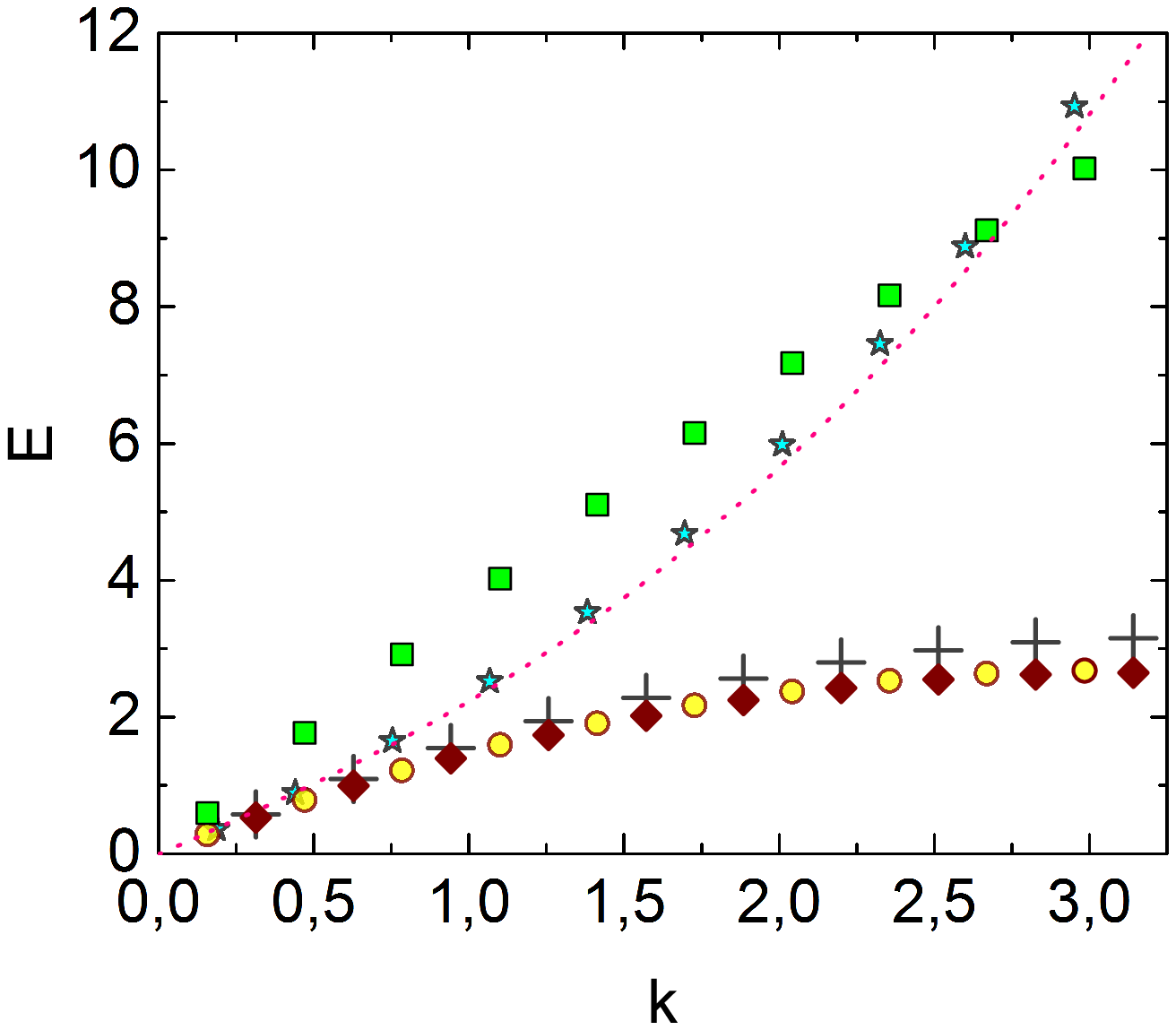} } \caption{
[Color online]  Dispersion laws $E(k)$ for quasiparticles of several
types for a 1D system of spinless point
bosons with zero boundary conditions and $N=1000$, $\bar{\rho}=1$, $\gamma =1$%
: 1)~Dispersion law for the states ($n_{j\leq N-1}=1$, $n_{N}=r>1$)
calculated using formulae (\ref{7-1}), (\ref{7-2}), and (\ref{13})
(stars); 2)~Bogoliubov law $E=\sqrt{k^{4}+4\gamma
\bar{\rho}^{2}k^{2}}$ (dotted
curve); 3) \textquotedblleft dispersion curve\textquotedblright\ $\frac{%
E(k=p/(j_{0}-1))}{j_{0}-1}$ for a condensate of quasiparticles for
$j_{0}=2$ (Lieb's hole excitations, diamonds), $10$ (circles), $100$ (crosses), and $%
1000 $ (squares); the condensate of quasiparticles is characterized
by the quantum numbers $n_{j\leq i}=1$, $n_{j>i}=j_{0}$ (in this
case, $N-i\gg 1$); see more details in the text.
 \label{fig9exek}}
\end{figure}

Let us consider this issue in detail using the example of our 1D system of
interacting spinless point bosons under zero BCs. Its ground state
corresponds to the quantum numbers $n_{j\leq N}=1$~\cite%
{gaudin1971,mt2015,mtjpa2017}. For excited states, any $n_{j}$ can
take the values $n_{j}=1,2,3,\ldots ,\infty
$~\cite{gaudin1971,mtjpa2017}. Consider the state  ($n_{j\leq
N-1}=1, n_{N}=r>1$). Its energy is higher than the energy $E_{0}$ of
the ground state by a value of
\begin{equation}
E(p)=\sum\limits_{j=1}^{N}(|\acute{k}_{j}|^{2}-|k_{j}|^{2}),  \label{7-1}
\end{equation}%
where $\{|\acute{k}_{j}|\}$ and $\{|k_{j}|\}$ are the solutions of Gaudin's
equations (\ref{13}) for the states $(n_{j\leq N-1}=1,n_{N}=r>1)$ and ($%
n_{j\leq N}=1$), respectively, and $p$ is the quasimomentum of a quasiparticle~%
\cite{mtsp2019}
\begin{equation}
p=\sum\limits_{j=1}^{N}(|\acute{k}_{j}|-|k_{j}|)-\frac{1}{L}%
\sum\limits_{l,j=1}^{N}\left( \arctan {\frac{c}{|\acute{k}_{l}|+|\acute{k}%
_{j}|}}-\arctan {\frac{c}{|k_{l}|+|k_{j}|}}\right) |_{j\neq l}=\frac{\pi
(r-1)}{L}.  \label{7-2}
\end{equation}%
Numerical calculations show that at $\gamma =1$ the curve $E(p)$ is
close to the Bogoliubov curve (see Fig.~\ref{fig9exek}), and at
$\gamma \ll 1$ it is extremely close to the Bogoliubov curve (not
shown in Fig.~\ref{fig9exek}).
Therefore, it is clear that at $\gamma \lesssim 1$ all states of the type $%
(n_{j\leq N-1}=1,n_{N}=r>1)$ correspond to the Bogoliubov
quasiparticles.

Now consider the states of the type $(n_{j\leq
N-2}=1,n_{N-1}=r_{2},n_{N}=r_{1})$ where $\left( r_{1},r_{2}\right)
>1$. Similarly to the analysis above, it can be shown that the
energy of
such states is equal to $E_{0}+E(p_{1})+E(p_{2})+\delta E_{2}$, where $%
p_{1}=\pi (r_{1}-1)/L$, $p_{2}=\pi (r_{2}-1)/L$, and $\delta E_{2}$
is a small correction (always $\delta E_{2}<0$ and $|\delta
E_{2}|\lesssim \lbrack E(p_{1})+E(p_{2})]/N$). It is natural to
interpret $\delta E_{2}$ as the interaction energy of two
quasiparticles corresponding to the states ($n_{j\leq N-1}=1,n_{N}=r_{1})$ and $%
(n_{j\leq N-1}=1,n_{N}=r_{2})$ of the system. Further, we can show that the state $%
(n_{j\leq N-3}=1,n_{N-2}=r_{3},n_{N-1}=r_{2},n_{N}=r_{1})$ with
$\left( r_{1},r_{2},r_{3}\right) >1$ can be considered as three
interacting quasiparticles corresponding to the states $(n_{j\leq
N-1}=1,n_{N}=r_{1})$, $(n_{j\leq N-1}=1,n_{N}=r_{2})$, and
$(n_{j\leq N-1}=1,n_{N}=r_{3})$. Whence it is not difficult to guess
that any state of the type $(n_{j\leq N-i}=1,n_{N-i+1}=r_{i},\ldots
,n_{N}=r_{1})$ is a set of $i$~interacting quasiparticles, where the
first quasiparticle corresponds to the state $(n_{j\leq
N-1}=1,n_{N}=r_{1})$, the second to the state $(n_{j\leq N-1}=1,n_{N}=r_{2})$%
, and so on, and the $i$-th particle to the state $(n_{j\leq
N-1}=1,n_{N}=r_{i})$. If $i\sim N$, then the interaction energy
$\delta E_{i}$ of
quasiparticles is high and comparable to the energy $%
E(p_{1})+\ldots +E(p_{i})$. The properties described above are valid
for any $\gamma >0$. Therefore, the state  $n_{j\leq N}=j_{0}>1$
corresponds to a condensate of $N$ identical interacting
quasiparticles, each of which corresponds to the state $(n_{j\leq
N-1}=1,n_{N}=j_{0})$.

Is the quasiparticle  $(n_{j\leq N-1}=1, n_{N}=j_{0}>1)$ an
elementary excitation? This question is not quite trivial. There are
two ways to answer it: (1)~to explore the structure of the wave
functions (WFs) or (2)~to diagonalize the Hamiltonian. Let us
consider these possibilities.

(1) For periodic BCs, \textit{any} excited state of a system of
spinless bosons with the momentum $\hbar \mathbf{p}$ can be
described by the exact wave function \cite{holes2020,yuv2}
\begin{equation}
\Psi _{\mathbf{p}}(\mathbf{r}_{1},\ldots ,\mathbf{r}_{N})=A_{\mathbf{p}}\psi
_{\mathbf{p}}\Psi _{0},  \label{7-3}
\end{equation}%
where%
\begin{eqnarray}
\psi _{\mathbf{p}} &=&b_{1}(\mathbf{p})\rho _{-\mathbf{p}}+\sum\limits_{%
\mathbf{q}_{1}\neq 0}^{\mathbf{q}_{1}+\mathbf{p}\neq 0}\frac{b_{2}(\mathbf{q}%
_{1};\mathbf{p})}{2!N^{1/2}}\rho _{\mathbf{q}_{1}}\rho _{-\mathbf{q}_{1}-%
\mathbf{p}}+\sum\limits_{\mathbf{q}_{1},\ textbf{q}_{2}\neq 0}^{\mathbf{q}%
_{1}+\mathbf{q}_{2}+\mathbf{p}\not=0}\frac{b_{3}(\mathbf{q}_{1},\mathbf{q}%
_{2};\mathbf{p})}{3!N}\rho _{\mathbf{q}_{1}}\rho _{\mathbf{q}_{2}}\rho _{-%
\mathbf{q}_{1}-\mathbf{q}_{2}-\mathbf{p}}+  \notag \\
&+&\ldots +\sum\limits_{\mathbf{q}_{1},\ldots ,\mathbf{q}_{N-1}\neq 0}^{%
\mathbf{q}_{1}+\ldots +\mathbf{q}_{N-1}+\mathbf{p}\not=0}\frac{b_{N}(\mathbf{%
q}_{1},\ldots ,\mathbf{q}_{N-1};\mathbf{p})}{N!N^{(N-1)/2}}\rho _{\mathbf{q}%
_{1}}\ldots \rho _{\mathbf{q}_{N-1}}\rho _{-\mathbf{q}_{1}-\ldots -\mathbf{q}%
_{N-1}-\mathbf{p}},  \label{7-4}
\end{eqnarray}%
$\Psi _{0}$ is the ground-state wave function, $\mathbf{r}_{j}$ denote the
atomic coordinates, $A_{\mathbf{p}}$ is a normalization constant, and $\rho
_{\mathbf{q}}=\frac{1}{\sqrt{N}}\sum_{j=1}^{N}e^{-i\mathbf{q}\mathbf{r}_{j}}$
are collective variables. The values of the coefficients $b_{j}(\mathbf{p})$
show which state is described by function (\ref{7-3}): with one, two, or $j$
interacting elementary excitations~\cite{holes2020}. If this is one
elementary excitation, then $b_{j}(\mathbf{p})\sim 1$ for all $j=1,\ldots ,N$%
. In the case of $l$ excitations with the total momentum $\hbar \mathbf{p}%
=\hbar \mathbf{p}_{1}+\ldots +\hbar \mathbf{p}_{l}$, one needs to make the
following changes in Eqs.~(\ref{7-3}) and (\ref{7-4}): $\Psi _{\mathbf{p}%
}\rightarrow \Psi _{\mathbf{p}_{1}\ldots \mathbf{p}_{l}}$, $\psi _{\mathbf{p}%
}\rightarrow \psi _{\mathbf{p}_{1}\ldots \mathbf{p}_{l}}$, $A_{\mathbf{p}%
}\rightarrow A_{\mathbf{p}_{1}\ldots \mathbf{p}_{l}}$, and $b_{j}(\mathbf{q}%
_{1},\ldots ,\mathbf{q}_{j-1};\mathbf{p})\rightarrow b_{j}(\mathbf{q}%
_{1},\ldots ,\mathbf{q}_{j-1};\mathbf{p}_{1},\ldots ,\mathbf{p}_{l},N)$ for
all $j$. A solution in the case of two excitations ($l=2$) with the momenta $%
\hbar \mathbf{p}_{1}$ and $\hbar \mathbf{p}_{2}$ was found in paper~\cite%
{holes2020}; in this case the following relationships have to be obeyed: $%
b_{j\geq 3}\sim 1$, $b_{1}(\mathbf{p}_{1},\mathbf{p}_{2},N)\sim N^{-1/2}$, $%
b_{2}(\mathbf{q}_{1};\mathbf{p}_{1},\mathbf{p}_{2},N)\sim N^{-1/2}$ for $%
\mathbf{q}_{1}\neq -\mathbf{p}_{1},-\mathbf{p}_{2}$, and $b_{2}(\mathbf{q}%
_{1};\mathbf{p}_{1},\mathbf{p}_{2},N)\sim N^{1/2}$ for $\mathbf{q}_{1}=-%
\mathbf{p}_{1},-\mathbf{p}_{2}$. The case $l>2$ can be considered
analogously~\cite{holes2020}. Thus, the structure of WFs (\ref{7-3})
uniquely shows how many elementary excitations a given state
contains. According to the analysis in works~\cite{holes2020,yuv2},
for a periodic system of spinless bosons, the elementary excitations
are the Bogoliubov quasiparticles.

For zero BCs no solution of kind (\ref{7-3}), (\ref{7-4}) has been
found.

(2) If the Hamiltonian of a system of spinless bosons is
diagonalized and the creation and annihilation operators of
quasiparticles satisfy the boson commutation relations, then such
quasiparticles are elementary excitations. Indeed, any composite
excitation must correspond to a creation
operator of the form $\hat{B}_{\mathbf{k}}^{+}=\hat{b}_{\mathbf{k}%
_{1}}^{+}\cdots \hat{b}_{\mathbf{k}_{j}}^{+}$, where $\hat{b}_{\mathbf{k}%
_{1}}^{+},\ldots ,\hat{b}_{\mathbf{k}_{j}}^{+}$ are the creation operators
of elementary excitations~$1,2,\ldots j$ that the composite excitation
consists of. If the relation $\hat{b}_{\mathbf{k}_{i}}\hat{b}_{\mathbf{k}%
_{i}}^{+}-\hat{b}_{\mathbf{k}_{i}}^{+}\hat{b}_{\mathbf{k}_{i}}=1$ $%
(i=1,2,\ldots j)$ is satisfied, then it is easy to verify that the
boson commutation relation $\hat{B}_{\mathbf{k}}\hat{B}_{\mathbf{k}}^{+}-\hat{B}_{%
\mathbf{\ k}}^{+}\hat{B}_{\mathbf{k}}=1$ is violated. Therefore, in
particular, the Bogoliubov quasiparticles are elementary
excitations.

The Hamiltonian of a 1D system of spinless bosons with zero BCs was
diagonalized in work~\cite{mtmethodbog}; in this case all
quasiparticles are the Bogoliubov ones. Since for a weak coupling
the dispersion law and the quasimomentum $p=\pi (r_{1}-1)/L$ of the
quasiparticles $(n_{j\leq N-1}=1, n_{N}=r_{1})$ considered above
coincide with the dispersion law and the quasimomentum of the
Bogoliubov quasiparticles, quasiparticles $(n_{j\leq N-1}=1,
n_{N}=r_{1})$ are \textit{elementary excitations}. In this case,
there are no other elementary excitations in the system because any
excited state $(n_{1},n_{2},\ldots ,n_{N})$ can be considered as a
set of interacting Bogoliubov quasiparticles. The state $n_{j\leq
N}=j_{0}>1$ is a set of $N$ identical interacting elementary
excitations $(n_{j\leq N-1}=1, n_{N}=j_{0})$.

It is generally accepted  that elementary excitations of a 1D system
of spinless \textit{point} bosons are \textquotedblleft
particles\textquotedblright\ (Bogoliubov quasiparticles) and
\textquotedblleft holes\textquotedblright \cite{lieb1963}. However,
this is not quite so: the analysis above shows that, for a weak
coupling, only Bogoliubov quasiparticles are elementary excitations
of such a system. In work~\cite{holes2020} it was shown in a
different way, that for a weak coupling and periodic BCs, \textit{a
hole is a composite quasiparticle}: it is a set of identical
interacting Bogoliubov phonons, each of which  has the momentum
$2\hbar \pi /L$.

In the Bethe-ansatz approach \cite{mt2015} and the Bogoliubov approach \cite%
{bog1947}, the energy of any weakly excited state of a system with
$N\gg 1$ is expressed by the formula
\begin{equation}
E=E_{0}+\sum\limits_{\mathbf{p}}N_{\mathbf{p}}E(\mathbf{p}),  \label{7-e}
\end{equation}%
where $E(\mathbf{p})$ is the energy of a Bogoliubov quasiparticle, and $N_{%
\mathbf{p}}$ is the number of quasiparticles. In this case, there is
a one-to-one correspondence between weakly excited states in the
Bethe-ansatz and Bogoliubov approaches. On the other hand, in the
Bethe-ansatz approach, each state of the system can be formally
considered  as a set of interacting holes~\cite{holes2020,lieb1963},
and a formula like (\ref{7-e}) can be written for the holes. Hence,
there exists a dualism between holes and Bogoliubov
quasiparticles~\cite{holes2020,lieb1963}. Therefore, in order to
find out which quasiparticles are elementary (indivisible)
excitations, a more subtle analysis is required.

Such an analysis for a periodic system was carried out in
Ref.~\cite{holes2020}, where the hole was as if zoomed in. In
particular,  a comparison was made between the solutions for holes
with the momenta $2\hbar \pi /L$ and $4\hbar \pi /L$, on the one
hand, and one- and two-phonon solutions obtained using WF
(\ref{7-3}), on the other hand. The structure of the corresponding
wave functions, the correction to the one-phonon energy, and the
magnitude of the two-phonon interaction energy clearly
indicate~\cite{holes2020} that for a weak coupling, a hole with the
lowest momentum ($2\hbar \pi /L$) is a phonon, whereas a hole with
the momentum $4\hbar \pi /L$ consists of two identical interacting
phonons. It is therefore obvious that a hole with the momentum
$j\cdot
2\hbar \pi /L$ is $j$ interacting identical phonons, each with the momentum $%
2\hbar \pi /L$. In other words, phonons (Bogoliubov quasiparticles)
are elementary excitations. For a strong coupling, the system
becomes fermion-like, a hole-like quasiparticle becomes similar to
the fermion hole, and the separation into \textquotedblleft
particles\textquotedblright\ and holes is justified. At the same
time, the properties of the system are strange: the structure of the
WFs shows that apparently only two quasiparticles, $(0,0,\ldots
,0,1)$ and $(-1,0,\ldots ,0.0)$, are elementary
excitations~\cite{holes2020}.

Under zero BCs, the hole-like quasiparticle corresponds to the
quantum numbers $n_{1\leq j\leq i}=1$, $n_{i<j\leq N}=2$ (where
$i=0,1,\ldots
,N-2 $ ) and the quasimomentum $p=\pi (N-i)/L$~\cite{mtsp2019}. Fig.~\ref%
{fig9exek} shows the dispersion law for hole-like quasiparticles,
which was obtained using formula (\ref{7-1}). The numbers $\{|\acute{k}%
_{j}|\}$ and $\{|k_{j}|\}$ were found by numerical solution of
Gaudin's equations (\ref{13}). Similarly to the case of a system
with periodic BCs,
under zero BCs a hole with the quantum numbers $n_{1\leq j\leq i}=1$, $%
n_{i<j\leq N}=2$ is a set of $N-i$ interacting phonons, each with
the quasimomentum $\pi /L$.

It is of interest to consider the states ($n_{j\leq
i}=1,n_{j>i}=j_{0})$, where $i=0,1,\ldots ,N-2$, which correspond to
the sets of $N-i$ interacting phonons, each of which has the
quasimomentum $(j_{0}-1)\pi /L$. Since the
total quasimomentum of the system is defined by the formula~\cite%
{mtsp2019}
\begin{equation*}
P(\{|k_{i}|\})=\sum\limits_{j=1}^{N}|k_{j}|-\frac{1}{L}\sum%
\limits_{l,j=1}^{N}\arctan {\frac{c}{|k_{l}|+|k_{j}|}}|_{j\neq l}=\frac{\pi
}{L}\sum\limits_{j=1}^{N}n_{j},
\end{equation*}%
the ground state corresponds to the quasimomentum $\pi N/L$, and the state $%
(n_{j\leq i}=1$, $n_{j>i}=j_{0})$ can effectively be considered as a single
\textquotedblleft excitation\textquotedblright\ with the quasimomentum $%
p=(N-i)(j_{0}-1)\pi /L$. Using formula (\ref{7-1}) it is easy to
numerically find the energy $E$ of such a \textquotedblleft
quasiparticle\textquotedblright . The dependence of the quantity
$E^{\prime }=\frac{E}{j_{0}-1}$ on $k=\frac{p}{j_{0}-1}$ for various
$j_{0}$'s is shown in Fig.~\ref{fig9exek}. One can see that if
$j_{0}\lesssim 100$, the explored dependence is close to the hole
dispersion law (at $j_{0}=2$ this dependence \textit{is} the hole
dispersion law). This curious property shows that a
\textquotedblleft dispersion law\textquotedblright\ similar to the
hole one can also be obtained for a set of identical quasiparticles,
each of which has the quasimomentum $(j_{0}-1)\pi /L$. The state
$n_{j\leq N}=j_{0}>1$ considered in the previous sections
corresponds to a set of $N$ such quasiparticles (the last point in
the curves $E^{\prime }(k)$ in Fig.~\ref{fig9exek}).  Note that the
deviation of the \textquotedblleft dispersion
curves\textquotedblright\ $E^{\prime }(k)$ as $k$ increases from the
asymptotes $E(k\rightarrow 0)= const \cdot k$ is associated with the
strengthening of the interaction between the quasiparticles forming
this state. Similarly, as $N$ increases, the total interaction of
atoms increases, which leads to the depletion of the condensate and
a decrease in the accuracy of the GP method. In this sense the
bending of the curve $E^{\prime
}(k)$ with increasing $k$ and the increase in the value of  $%
\frac{E_{\mathrm{GP}_{N}}-E_{\mathrm{Bethe}}}{E_{\mathrm{Bethe}}}$
with increasing $N$ (Fig.~\ref{fig10-16exeN}) are kindred.

In the previous sections we saw that the state  $n_{j\leq
N}=j_{0}>1$ corresponds to the profile $\rho (x)$ with $j_{0}$
domains. A region
between adjacent domains is similar to a \textit{soliton} \cite%
{pethick2008,agraval2003}: it is a black soliton in the GP
(GP$_{N}$) approach and a dark soliton in the Bethe-ansatz approach.
In particular, the hole quasiparticle with quantum numbers $n_{j\leq
N}=2$ corresponds to two domains and one soliton. Soliton-like
properties of hole quasiparticles were
studied in works~\cite%
{tsuzuki1971,syrwid2015,kulish1976,ishikawa1980,sato2012,sato2016,brand2018}%
, see also review \cite{syrwid2021}. In the work \cite{syrwid2017}
by A. Syrwid and K. Sacha (see also \cite{syrwid2021}), it was
already found that
the particle density profile for the state  $n_{j\leq N}=j_{0}>1$ has a $%
j_{0}$-domain structure, but a relationship with the solutions of
the GP equation has not been explored. Note that a hole is a
soliton-like structure only if its momentum (quasimomentum) is large
enough, i.e. if it is comparable to the maximum possible value. If
the hole momentum is small, then a hole is a set of a small number
of interacting phonons and is not a soliton:  It is obvious without
calculations that several identical phonons with the smallest
momentum (quasimomentum) in a system with a very large $N$
should produce a non-soliton density profile  very close to $%
\rho (x)=\mathrm{const}$. In this case, such a state is a Lieb's
hole.

Finally, note that the choice of quasiparticles determines the
statistics. The WFs of a Bose system are always Bose-symmetric with
respect to atomic permutations. But the energy distribution of
quasiparticles and the commutation relations for the operators of
creation and annihilation of quasiparticles depend on the method of
introducing quasiparticles. Since quasiparticles can be introduced
in an infinite number of ways, the number of statistics is also
infinite. As was remarked above, elementary quasiparticles for a
system of spinless bosons with weak coupling are Bogoliubov
quasiparticles. If the system is one-dimensional and the interaction
is point-like, the partition function can be summed exactly for
$N=\infty ,L=\infty $ and $T\rightarrow 0$, which leads to a formula
for the free energy of a system of \textit{free Bose
quasiparticles}, for any coupling constant $\gamma
>0$~\cite{mt2015}. In general, the simplest statistical description of a system of interacting
bosons at $T\rightarrow 0$ is obtained if the system is considered
as an ideal gas of elementary excitations. A more complicated
thermodynamic approach was proposed by C.N. Yang and C.P.
Yang~\cite{yangs1969}. This is a universal single-particle approach.
At $T\rightarrow 0$ this approach is equivalent to the
Bose-quasiparticle approach, as shown in Ref.~\cite{mt-therm}. Based
on the Yang--Yang approach, it was found that at any $T$, a system
of interacting point bosons can be statistically considered as a
system of free
particles with a generalized fractional statistics \cite%
{bernard1994,isakov1994} (see also~\cite{khare}).
It is essential that
the approach in Refs.~\cite{bernard1994,isakov1994} is
single-particle and has no relevance to collective excitations.

\section{Discussion}

Above we considered the states  $n_{j\leq N}=j_{0}>1$, which
correspond simultaneously to a condensate of atoms and a condensate
of elementary quasiparticles. Using the Bethe ansatz, it is easy to
write down
states with several condensates of quasiparticles. For example, the state $%
(n_{1}=n_{2}=\ldots =n_{j=10^{6}}=r_{1},n_{j=10^{6}+1}=\ldots
=n_{j=2\times 10^{6}}=r_{2},n_{j=2\times 10^{6}+1}=\ldots
=n_{j=3\times 10^{6}=N}=r_{3})$, where $r_{1},r_{2},r_{3}>1$,
$r_{1}\neq r_{2}, r_{3}$, and $r_{2}\neq  r_{3}$%
, corresponds to three condensates of quasiparticles (and three
condensates of atoms, if $\gamma $ is small). But such states can
scarcely be obtained experimentally and cannot be described by the
Gross-Pitaevskii equation.

Condensates of photons \cite{weitz2010}, excitons \cite%
{gorbunov2006,high2012,glazov2020}, exciton polaritons \cite%
{kasprzak2006,west2007,little2007,yamamoto2014}, and magnons \cite%
{borovik1984,demokritov2006,bunkov2010,dzyapko2010,bunkov2013} have
already been obtained experimentally (see also
review~\cite{ketterson2013}). The possibility of condensation of
rotons \cite{iordanskii1980} and phonon pairs \cite{kagan2007} was
argued.

In work~\cite{kagan2007} an effectively one-dimensional Bose gas was
considered, and it was shown that a periodic variation of the
transverse trap frequency can create a condensate of pairs of
phonons with opposite momenta $\mathbf{k}_{z},-\mathbf{k}_{z}$. This
is a \textquotedblleft doubly coherent\textquotedblright\ state: a
condensate of atoms and a condensate of phonon pairs. In this
approach, in contrast to our solution, an external field is required
for the condensate of quasiparticles to exist, pairs of
quasiparticles are condensed instead of single quasiparticles, and
the number of phonons in the condensate is much smaller than $N$
(this is evident from the fact that the condensate of quasiparticles
changes the particle density $\rho (z)$ very little).

\section{Concluding remarks}

We have shown that for a 1D system of spinless point bosons under
zero boundary conditions, each stationary excited state $\Phi
_{j_{0}}(x)$ of a condensate of $N$ atoms contains a condensate of
$N$ elementary excitations (Bogoliubov quasiparticles); we have
proved this for the case $N\leq 5$ and argued for $N>5$. It is
interesting that the solution $\Phi _{j_{0}}(x)$ has a
$j_{0}$-domain density profile $\rho (x)$, and the region between
neighbouring domains is a stationary black soliton (i.e. the solution contains $%
j_{0}-1$ identical black solitons).

Thus, any stationary excited state of a condensate of atoms is
doubly coherent: it is simultaneously a condensate of atoms and a
condensate of elementary excitations. From the other side, this
means that for a weak coupling every state with a condensate of $N$
elementary excitations corresponds to a condensate of atoms. We have
shown this for a 1D system with a point-like interatomic potential.
However, it is natural to expect that this property is universal and
holds true for a Bose system of any dimensionality, with any
interatomic potential and any trap field (or any BCs in the absence
of a trap).

Note that the solutions $\Phi _{j_{0}>1}(x)$ have already been
obtained in the form of elliptic functions~\cite{carr2000r}, and the
soliton-like profile $\rho (x)$ for the Bethe-ansatz states
$n_{j\leq N}=j_{0}>1$ has previously been found
in~\cite{syrwid2017}. However, a relation of these solutions to each
other and to elementary quasiparticles has not been ascertained.
This was done in the present work. It is also worth noting that the
soliton-like behavior of the solutions $\Phi _{j_{0}>1}(x)$ is due
to the condensate of elementary excitations. This can be verified by
finding the particle density profile $\rho (x)$ within the
Bethe-ansatz approach for a system with one, two, ten (say) and $N$
identical elementary excitations. The change in the profile with
increasing number of identical excitations should show that the
solitonic structure appears when the number of identical excitations
becomes large [$\sim N$] (it is clear that for $N\gg 1$ several
excitations give a non-soliton profile $\rho(x)$ which, far from the
boundaries, is close to $\rho(x)= const$). This tendency can be seen
even for $N=3$ (see Fig.~\ref{fig18exroN3}, panel for $j_{0}=3$). It
is clear that this statement should be true for large $N$, since
many identical elementary excitations resonantly amplify the
deviation of $\rho (x)$ from the constant. In this case the largest
deviation should occur at the points corresponding to the
oscillation maximum for each half-wave $\lambda /2=\pi
/p=L/(j_{0}-1)$, which exactly corresponds to the structure of $\Phi
_{j_{0}>1}(x)$. Thus, a $(j_{0}-1)$-fold dark soliton is a $N$-fold
amplified phonon with the quasimomentum $\hbar \pi (j_{0}-1)/L$. It
agrees with the result of work~\cite{kagan2007}, according to which
the condensate of phonon pairs produces a periodic particle density
profile.

The doubly coherent states are soliton-like and can have a long
lifetime. One can try to obtain them using an external alternating
electromagnetic field in two ways: abruptly or gradually. In the
first case, the field transforms the condensate, as a set of atoms,
into the
excited state $\Phi _{j_{0}}(x)$, and the field frequency must be equal to $%
(E_{j_{0}}-E_{1})/\hbar $, where $E_{j_{0}}$ and $E_{1}$ are the
condensate energies for the states $\Phi _{j_{0}}(x)$ and $\Phi
_{1}(x)$, respectively. In the second case, the external field acts
on the system,  as a set of quasiparticles, and gradually creates
identical quasiparticles;  the energy of the field quantum should be
equal to the quasiparticle energy. In this case, because the number
of identical quasiparticles increases and the energy of each
quasiparticle decreases due to its interaction with other
quasiparticles, the field frequency must be correspondingly lowered
in time.  The derivation \cite{lt2} of the probability of the
creation of a circular roton in He~II by the field of a microwave
resonator~\cite{svh1,svh2} is related to the second mechanism.

It is clear that an experimentally obtained condensate of elementary
quasiparticles will contain an admixture of other quasiparticles.
The criterion for the presence of a condensate of elementary
excitations can be a strong  nonuniformity of the particle density
profile $\rho (\mathbf{r})$ that is not associated with the trap;
because a large number of identical elementary
excitations produce a nonuniform particle density profile $\rho (\mathbf{r})$%
, whereas a large number of various elementary excitations produce a
uniform profile $\rho (\mathbf{r})\approx const$. Another
possibility for the experimental creation of a stationary excited
condensate state was  proposed in work~\cite{yukalov1997}%
. The idea of parametric resonance was explored in~\cite{kagan2007}.

We believe
that doubly coherent states of different symmetries will be obtained
experimentally in the future. It is not excluded that they can  be
obtained for He~II as well.


\section*{Acknowledgements}

The author gratefully acknowledges the National Academy of Sciences of
Ukraine and the Simons Foundation for financial support.

\section*{Appendix. Is the condensate fragmented?}

According to formula (\ref{4-1}) and the solutions for the coefficients $%
g_{l}$, the condensate $\Phi _{j_{0}}(x)$ can disperse into many
$l$-harmonics. Let us see whether such a condensate is fragmented.

It is known that the structure of a condensate for the state $\Psi
(x_{1},\ldots ,x_{N})$ is determined using the diagonal expansion of
the single-particle density matrix
\begin{equation}
F_{1}(x,x^{\prime })=\sum \limits_{j=1}^{\infty }\lambda _{j}\phi _{j}^{\ast
}(x^{\prime })\phi _{j}(x).  \label{6-1}
\end{equation}%
A fragmented condensate corresponds to the case when $\lambda
_{j}\sim N$ for two (or more) $j$'s in~(\ref{6-1}). Let us assume
that the condensate is fragmented into two ones (1 and 2) such that
$\lambda _{1}\sim \lambda _{2}\sim N$ and $\lambda _{j>2}=0$. The
wave function of such a system looks like
\begin{equation}
\Psi _{2c}(x_{1},\ldots ,x_{N},t)=\mathrm{const}\,\sum\limits_{P}\prod%
\limits_{j=1}^{\lambda _{1}}\ \psi _{1}(x_{j},t)\prod\limits_{l=\lambda
_{1}+1}^{N}\psi _{2}(x_{l},t),  \label{6-0}
\end{equation}%
where summation over the permutations ($P$) $x_{j}\leftrightarrow x_{l}$
provides the Bose symmetry of $\Psi _{2c}$. Substituting function (\ref{6-0}%
) into the Schr\"{o}dinger equation gives an equation for two functions, $%
\psi _{1}(x,t)$ and $\psi _{2}(x,t)$, instead of the GP equation.
Therefore, the GP and GP$_{N}$ equations describe a non-fragmented
condensate. It can be shown that if expansion (\ref{6-1}) contains
$\lambda _{1}=N$ and $\lambda _{j>1}=0$, then the system is
described by a single-condensate wave function
\begin{equation}
\Psi (x_{1},\ldots ,x_{N},t)=e^{Et/i\hbar }\prod\limits_{j=1}^{N}\psi (x_{j})
\label{p3}
\end{equation}%
with $\psi (x)=\phi _{1}(x)$.

For the operator approach, the picture is less obvious. Let the
density matrix (\ref{6-1}) have the macroscopically occupied
orbitals $\phi _{1}(x)$ and $\phi _{2}(x)$, and let $N\gg 1$. Then,
in the expansion
\begin{equation}
\hat{\psi}(x,t)=\sum_{j=1,2,\ldots ,\infty }\hat{d}_{j}(t)\phi _{j}(x),
\label{6-00}
\end{equation}%
$\hat{d}_{1}(t)$ and $\hat{d}_{2}(t)$ can be regarded as
$c$-numbers, i.e. $\hat{d}_{1}(t)=d_{1}(t)$ and
$\hat{d}_{2}(t)=d_{2}(t)$. In this case,
the second-quantized operator $\hat{\psi}(x,t)$ can be represented as $%
\hat{\psi}(x,t)=\Psi (x,t)+\hat{\vartheta}(x,t)$, where $\Psi
(x,t)=d_{1}(t)\phi _{1}(x)+d_{2}(t)\phi _{2}(x)$ is the condensate wave
function, and $\hat{\vartheta}(x,t)$ is a small operator correction. If we
put $\hat{\vartheta}(x,t)=0$, then the Heisenberg equation for $\hat{\psi}%
(x,t)$ becomes a time-dependent Gross equation for $\Psi (x,t)$, and
the density matrix is given by the formula
\begin{equation}
F_{1}(x,x^{\prime })\equiv \langle \hat{\psi}^{+}(x^{\prime },t)\hat{\psi}%
(x,t)\rangle =\Psi ^{\ast }(x^{\prime },t)\Psi (x,t)=\Phi ^{\ast }(x^{\prime
})\Phi (x),  \label{6-2}
\end{equation}%
where we pass to the stationary solution $\Psi (x,t)=e^{\epsilon
t/i\hbar
}\Phi (x)$. Formula (\ref{6-2}) coincides with expansion (\ref{6-1}%
) if $\lambda _{1}=N$, $\lambda _{j\geq 2}=0$, $\phi _{1}(x)=\Phi (x)/\sqrt{N%
}$, and $\phi _{j\geq 2}(x)$ are some functions orthogonal to $\phi _{1}(x)$%
. That is, we obtained that $\Phi (x)$ describes a non-fragmented
condensate, although we proceeded from a fragmented condensate
($\lambda _{1}\sim \lambda _{2}\sim N$). Later we will return to
this issue.

The formulae $\hat{\Psi}(x,t)=\hat{a}_{0}\Psi (x,t)/\sqrt{N}$ and
$\langle \hat{a}_{0}^{+}\hat{a}_{0}\rangle =N$ also lead to
Eq.~(\ref{6-2}). Let us show that formula (\ref{6-2}) with $\Phi
(x)=\Phi _{j_{0}}(x)$ [see Eq.~(\ref{4-1})] necessarily implies a
non-fragmented condensate according to the criterion based on
formula~(\ref{6-1}). This conclusion already follows from the fact
that the diagonal expansion (\ref{6-1}) is unique; therefore, it is
impossible to express $F_{1}(x,x^{\prime })$ as a different
expansion of type~(\ref{6-1}). Let us prove this statement by
contradiction. Assume that besides expansion (\ref{6-2}), there is
another expansion~(\ref{6-1}). It is convenient to pass from
Eq.~(\ref{6-1}) to the equivalent system of equations
\begin{equation}
\int\limits_{0}^{L}dx^{\prime }\phi _{j}(x^{\prime })F_{1}(x,x^{\prime
})=\lambda _{j}\phi _{j}(x),\quad j=1,2,\ldots ,\infty .  \label{6-4}
\end{equation}%
From formulae (\ref{6-1}) and (\ref{6-2}) [with $\Phi (x)=\Phi _{j_{0}}(x)$%
], it is clear that we may seek the functions $\phi _{j}(x)$ from Eq.~(\ref%
{6-1}) in the form
\begin{equation}
\phi _{j}(x)=\sum\limits_{l=1,2,\ldots ,\infty }A_{l}^{(j)}\sqrt{2/L}\cdot
\sin {[\ pij_{0}(2l-1)x/L]},  \label{6-5}
\end{equation}%
which is similar to expansion~(\ref{4-1}). Let us substitute
(\ref{6-5}) in (\ref{6-4}) and take Eq. (\ref{6-2}), where
$\Phi(x)=\Phi_{j_{0}}(x)$, and Eq. (\ref{4-1}) into account. After
simple algebra, we find the equation
\begin{equation*}
N\sum\limits_{p,l=1,2,\ldots ,\infty }A_{l}^{(j)}g_{l}g_{p}\sin {[\pi
j_{0}(2p-1)x/L]}=\lambda _{j}\sum\limits_{p=1,2,\ldots ,\infty
}A_{p}^{(j)}\sin {[\pi j_{0}(2p-1)x/L]}.
\end{equation*}%
Since the functions $\sin {[\pi j_{0}(2p-1)x/L]}$ are independent, we equate
the coefficients of the functions $\sin {[\pi j_{0}(2p-1)x/L]}$ to zero and
get the system of equations for the coefficients $A_{l}^{(j)}$ for each $%
j=1,2,\ldots ,\infty $:
\begin{equation}
\sum\limits_{l=1,2,\ldots ,\infty }A_{l}^{(j)}g_{l}=\frac{\lambda _{j}}{N}%
\frac{A_{1}^{(j)}}{g_{1}}=\frac{\lambda _{j}}{N}\frac{A_{2}^{(j)}}{g_{2}}%
=\ldots =\frac{\lambda _{j}}{N}\frac{A_{\infty }^{(j)}}{g_{\infty }}.
\label{6-7}
\end{equation}%
Let $j=1$ and $\lambda _{1}\neq 0$. Then from Eq.~(\ref{6-7}) we find $%
A_{p}^{(1)}=\frac{g_{p}}{g_{1}}A_{1}^{(1)}$, where $p=1,2,\ldots
,\infty $. From formulae (\ref{4-1}),
$A_{l}^{(1)}=\frac{g_{l}}{g_{1}}A_{1}^{(1)}$ and (\ref{6-5}) with
$j=1$, it follows that $\phi
_{1}(x)=(A_{1}^{(1)}/g_{1})\Phi _{j_{0}}(x)/\sqrt{N}$. Using the equations $%
\int\limits_{0}^{L}dx|\Phi _{j_{0}}(x)|^{2}=N$ and $\int_{0}^{L}dx\phi
_{1}^{\ast }(x)\phi _{1}(x)=1$, we get $A_{1}^{(1)}=\pm g_{1}$, whence $\phi
_{1}(x)=\Phi _{j_{0}}(x)/\sqrt{N}\equiv \Phi (x)/\sqrt{N}$. Substituting $%
A_{l}^{(1)}=\frac{g_{l}}{g_{1}}A_{1}^{(1)}$ into Eq.~(\ref{6-7}) with $j=1$,
we obtain
\begin{equation}
\lambda _{1}/N=\sum\limits_{l=1,2,\ldots ,\infty }g_{l}^{2}.  \label{6-8}
\end{equation}%
On the other hand, normalization condition (52) from work~\cite{gp1} gives $%
\sum_{l=1,2,\ldots ,\infty }g_{l}^{2}=1$. Hence, $\lambda _{1}=N$. Further, $%
\int_{0}^{L}dxF_{1}(x,x)=\int\limits_{0}^{L}dx|\Phi (x)|^{2}=N$, and from
Eq.~(\ref{6-1}) it follows that $\int\limits_{0}^{L}dxF_{1}(x,x)=\lambda
_{1}+\lambda _{2}+\ldots +\lambda _{\infty }$, where $\lambda _{j}\geq 0$
for any $j$. As a result, we obtain $\lambda _{j\geq 2}=0$. According to
Eq.~(\ref{6-7}), if $\lambda _{j\geq 2}=0$, then $\sum_{l=1,\ldots ,\infty
}A_{l}^{(j)}g_{l}=0$ must hold for $j\geq 2$. However, it is easy to see
that this equation is equivalent to the orthogonality condition $%
\int_{0}^{L}dx\phi _{j\geq 2}^{\ast }(x)\phi _{1}(x)=0$. So we have shown
that the density matrix (\ref{6-2}) corresponds to the diagonal expansion (%
\ref{6-1}) with $\lambda _{1}=N$, $\lambda _{j\geq 2}=0$, and $\phi
_{1}(x)=\Phi (x)/\sqrt{N}=\Phi _{j_{0}}(x)/\sqrt{N}$. Furthermore, the
functions $\phi _{j\geq 2}(x)$ are orthogonal to $\phi _{1}(x)$.

Thus, the wave function $\Phi _{j_{0}}(x)$ (\ref{4-1}) describes a
non-fragmented condensate for any  values of the parameters and any
$j_{0}=1,2,\ldots ,\infty $.

It was noted above that if we proceed from a fragmented condensate with $%
\lambda _{1}\sim \lambda _{2}\sim N$, then the approximation $\hat{d}%
_{1}(t)=d_{1}(t)$ and $\hat{d}_{2}(t)=d_{2}(t)$ leads to a
non-fragmented condensate. This means that it is incorrect to
describe the fragmented condensate by the formulae
$\hat{d}_{1}(t)=d_{1}(t)$, $\hat{d}_{2}(t)=d_{2}(t)$. Only one
operator, say $\hat{d}_{1}(t)$, can
be considered as a c-number: $\hat{d}_{1}(t)=d_{1}(t)$. Then $\hat{%
d}_{2}(t)$ must be considered as an operator. In this case it is
difficult to diagonalize the Hamiltonian, because corrections such as $\hat{d}%
_{2}^{3}$ and $\hat{d}_{2}^{4}$ have to be taken into account. But if $%
\lambda _{1}\gg \lambda _{2}\gg 1$, it is sufficient  to consider only $\hat{d}%
_{2}^{2}$, and the Hamiltonian can be diagonalized like as in
Bogoliubov's method. A somewhat more complicated approach has shown
that a condensate in a 1D weakly interacting Bose gas under zero BCs
and at
zero temperature can be fragmented~\cite{mtfragm2}. The results of work~\cite%
{mtfragm2} are consistent with the solution obtained for periodic
BCs by a different method~\cite{mtfragm1}.

\renewcommand\refname{}


\end{document}